%% file: main.tex
\documentclass[twocolumn]{aastex7}

\usepackage{threeparttable} 
\usepackage{graphicx}      
\usepackage{xcolor}   
\usepackage{xspace}
\usepackage{booktabs}
\usepackage{subcaption}
\usepackage{hyperref}
\usepackage{wrapfig}
\usepackage{sidecap}
\usepackage{float}
\usepackage{placeins}
\usepackage{capt-of} 
\usepackage{amssymb, amsmath, mathtools}
\usepackage{natbib}%, authblk}
\usepackage{enumerate}

\usepackage{latexsym}
\usepackage[utf8]{inputenc}
\usepackage[T1]{fontenc}
\usepackage[version=4]{mhchem} % Molecular species! e.g. \ce{H2O}
\usepackage{xspace}

\newcommand{\eureka}{\texttt{Eureka!}\xspace}
\newcommand{\exotedrf}{\texttt{exoTEDRF}\xspace}
\newcommand{\planet}{LTT~9779~b}
\newcommand{\planetname}{\planet} %\xspace}

\input{kbsmacs.tex}
\graphicspath{{./}{Figures/}}

\begin{document}

\onecolumngrid

% ---- Hard-coded title for arXiv ----
\begin{center}
    \Large\textbf{Heat Reveals What Clouds Conceal: \\ Global Carbon \& Longitudinally Asymmetric Chemistry on LTT~9779~b}
\end{center}

% ---- Hard-coded authors ----
\begin{center}
    Reza Ashtari$^{1}$, Sean Collins$^{2,3,4}$, Jared Splinter$^{5}$, Kevin B. Stevenson$^{1}$, Vivien Parmentier$^{6}$, \\
    Jonathan Brande$^{7,8}$, Suman Saha$^{9,10}$, Sarah Stamer$^{11}$, Ian J.\ M.\ Crossfield$^{7}$, James S.\ Jenkins$^{9,10}$, \\
    K.\ Angelique Kahle$^{12,13}$, Joshua D.\ Lothringer$^{14}$, Nishil Mehta$^{6,15}$, Nicolas B.\ Cowan$^{4,5}$, \\
    Diana Dragomir$^{11}$, Laura Kreidberg$^{12}$, Thomas M.\ Evans\textendash Soma$^{16}$, Tansu Daylan$^{17}$, \\
    Olivia Venot$^{18}$, Xi Zhang$^{19}$
\end{center}

% ---- Hard-coded affiliations (no street addresses) ----
\begin{center}
\textit{
$^{1}$Johns Hopkins University Applied Physics Laboratory, Laurel, MD, USA\\
$^{2}$University of British Columbia, Vancouver, BC, Canada\\
$^{3}$Institut Trottier de recherche sur les exoplanètes (iREx), Université de Montréal, Montréal, QC, Canada\\
$^{4}$Department of Physics, McGill University, Montréal, QC, Canada\\
$^{5}$Department of Earth \& Planetary Sciences, McGill University, Montréal, QC, Canada\\
$^{6}$Université Côte d'Azur, Observatoire de la Côte d'Azur, CNRS, Laboratoire Lagrange, Nice, France\\
$^{7}$The University of Kansas, Lawrence, KS, USA\\
$^{8}$University of Maryland, College Park, MD, USA\\
$^{9}$Universidad Diego Portales, Santiago, Chile\\
$^{10}$Centro de Excelencia en Astrofísica y Tecnologías Afines (CATA), Santiago, Chile\\
$^{11}$University of New Mexico, Albuquerque, NM, USA\\
$^{12}$Max Planck Institute for Astronomy, Heidelberg, Germany\\
$^{13}$Department of Physics and Astronomy, Heidelberg University, Heidelberg, Germany\\
$^{14}$Space Telescope Science Institute, Baltimore, MD, USA\\
$^{15}$Université Côte d'Azur, Laboratoire Lagrange, Nice, France\\
$^{16}$School of Information and Physical Sciences, University of Newcastle, Callaghan, NSW, Australia\\
$^{17}$Department of Physics, Washington University in St.\ Louis, St.\ Louis, MO, USA\\
$^{18}$Université Paris Cité and Université Paris-Est Créteil, CNRS, LISA, Paris, France\\
$^{19}$Department of Earth and Planetary Sciences, University of California, Santa Cruz, Santa Cruz, CA, USA
}
\end{center}

\vspace{0.2cm}
\begin{center}
    \textit{Submitted to AJ}
\end{center}

% \vspace{-0.5cm}

\begin{abstract}

\planetname{} is an ultra-hot Neptune ($R_p \approx 4.7\,R_\oplus$, $M_p \approx 29\,M_\oplus$) orbiting its Sun-like host star in just 19 hours, placing it deep within the ``hot Neptune desert,'' where Neptunian planets are seldom found. We present new JWST NIRSpec G395H phase-curve observations that probe its atmospheric composition in unprecedented detail. At near-infrared wavelengths, which penetrate the high-altitude clouds inferred from previous NIRISS/SOSS spectra, thermal emission reveals a carbon-rich atmosphere with opacity dominated by carbon monoxide (CO) and carbon dioxide (CO$_2$). Both species are detected at all orbital phases, with retrieved mixing ratios of $\sim$$10^{-1}$ for CO and $\sim$$10^{-4}$ for CO$_2$, indicating a globally well-mixed reservoir of carbon-bearing gases. We also moderately detect water vapor (H$_2$O) and tentatively detect sulfur dioxide (SO$_2$), providing insight into its chemistry and possible photochemical production under intense stellar irradiation. From these detections we infer a carbon-to-oxygen ratio near unity (C/O~$\approx$~1) and a metallicity exceeding $500\times$ Solar, consistent with equilibrium chemistry predictions for high-temperature atmospheres. This enrichment raises the mean molecular weight, reducing atmospheric escape, and likely helps \planetname{} retain a substantial atmosphere despite extreme irradiation. Our findings show that \planetname{} survives where few planets can, maintaining a carbon-rich atmosphere in a region where hot Neptune-class worlds are expected to evaporate. This makes \planetname{} a valuable laboratory for studying atmospheric escape and chemical processes under extreme conditions, offering new insight into the survival of planets in the hot Neptune desert.
\\
% The presence of SO$_2$ alongside H$_2$O suggests sulfur photochemistry at the limb, where material transported from the dayside is exposed to strong ultraviolet flux, echoing detections in other highly irradiated gas giants. 

\end{abstract}

%% Keywords should appear after the \end{abstract} command. 
%% See the online documentation for the full list of available subject
%% keywords and the rules for their use.

\clearpage
\section{Introduction}
\label{sec:intro}

% Show in Earth or Jupiter Units? 0.0923Mj, 0.4211Rj

Ultra-hot Neptunes are a rare class of exoplanets occupying the ``Neptune desert'', a demographic of Neptune-sized planets at very short orbital periods \citep{Mazeh2016}. The discovery of \planetname{} in this regime (orbital period $\sim$0.79 days, radius $\sim$4.7 $R_\oplus$) provided the first-known exoplanet of this kind \citep{Jenkins2020}. With an equilibrium temperature around 2000 K, \planetname{} is the most highly irradiated Neptune-size planet known. Straddling the equilibrium temperature boundary between hot Jupiters and Ultra-Hot Jupiters, it resides in the demographic region where a planet’s near-infrared emission begins to dominate over reflected optical light \citep{Jenkins2020}. Its existence, a Neptune-mass ($\sim$29 $M_\oplus$) planet so close to its star, challenges models of atmospheric evaporation and planetary formation, making it a compelling target to understand how such a planet can retain a substantial atmosphere in extreme conditions \citep{Lopez2017,Jenkins2020,Radica2025}. 

 Using Spitzer IRAC at 3.6 and 4.5 $\mu$m, \citet{Crossfield2020} and \citet{Dragomir2020} measured the planet’s dayside emission and day-night contrast. The dayside brightness temperature was found to be on the order of 2100 K, while the nightside was $\sim$1100 K cooler,  indicating very inefficient heat transport from day to night \citep{Crossfield2020}. A minimal phase offset of $-10^{\circ} \pm 21^{\circ}$ eastward was detected, with the planet’s maximum emission near the substellar point \citep{Crossfield2020}. Intriguingly, the planet’s 4.5 $\mu$m flux was lower than expected from a blackbody, suggesting the presence of strong molecular absorption by CO and/or CO$_2$, implying a non-inverted atmosphere, in contrast with the ultra-hot Jupiter population. \citep{Dragomir2020}. In fact, \planetname’s Spitzer eclipse spectrum provided the first evidence of CO/CO$_2$ absorption in a Neptune-size exoplanet, pointing to a high metallicity composition \citep{Dragomir2020}. The dayside temperature was cooler than predicted for a zero-albedo planet, suggesting a significant fraction of the stellar irradiation is reflected rather than absorbed \citep{Crossfield2020}. Together, these early results implied that \planetname{} might harbor reflective, high-altitude clouds, and an atmosphere enriched in heavy elements (as expected for a Neptune-mass planet). 

Subsequent observations in the optical confirmed the exceptionally high albedo of \planetname’s dayside \citep{Hoyer2023,Saha2025}. Using the CHEOPS space telescope, \citet{Saha2025} detected the secondary eclipses of \planetname{} at visible wavelengths and measured a geometric albedo of $A_g $\sim$0.73\pm0.11$,  comparable to Venus and far higher than typical hot Jupiters (which often have $A_g < 0.1$). This makes \planetname{} one of the brightest exoplanets on record \citep{Hoyer2023,Saha2025}. Modeling by Saha et al. indicated that such a high reflectance, combined with the Spitzer infrared fluxes, requires an extremely metal-rich atmosphere (on the order of hundreds of times solar metallicity) and the presence of reflective clouds on the dayside \citep{Saha2025}. Indeed, a lower limit of $\sim$200$\times$ solar metallicity was inferred, along with cloud condensates likely composed of silicate material \citep{Hoyer2023, Saha2025}. This finding was then reinforced by HST/WFC3 ultraviolet-visible observations. 

Although HST could not conclusively detect the planet’s UV/blue eclipse (yielding an upper limit of $\sim$113ppm on the 0.2-0.8 $\mu$m flux), the data constrained the planet’s reflectance at those wavelengths \citep{Radica2025}. Comparison to cloud models showed that silicate condensates (e.g. Mg-silicates) with small, highly reflective particles best explain \planetname’s large albedo and flat reflectance spectrum \citep{Radica2025}. In other words, the planet likely hosts metallic silicate clouds that scatter a substantial portion of incoming light, thereby cooling the dayside. This reflective shield may even help the planet retain its atmosphere by reducing stellar heating, as speculated by \citet{Radica2025}. 

Thanks to the James Webb Space Telescope (JWST), our understanding of \planetname’s atmosphere has advanced dramatically. Recently, JWST/NIRISS SOSS observations (0.6-2.8$\mu$m) captured a full phase curve and transmission spectrum of \planetname{} \citep{Radica2024, Coulombe2025}. The transmission spectrum obtained during the planet’s transit showed only muted spectral features,  significantly flatter than a clear solar-composition atmosphere would produce \citep{Radica2024}. This ruled out a featureless spectrum at $>5\sigma$ confidence, yet the observed spectral amplitude could be explained by a broad range of high metallicities (20-850× solar) in combination with clouds/hazes \citep{Coulombe2025}. Retrieval analyses favor a scenario where \planetname’s terminator region has clouds at millibar pressures and a heavy mean molecular weight, consistent with a water- or methane-dominated composition at high metallicity \citep{Coulombe2025}. Notably, the temperature-profile and condensation models indicate that silicate clouds can condense at the terminator of \planetname{} \citep{Coulombe2025}. These clouds, formed near the planet’s morning limb, could be carried by eastward winds onto the dayside, thereby contributing to the brilliant reflective clouds observed and perhaps creating a feedback mechanism that helps the planet avoid complete atmospheric loss \citep{Coulombe2025}. In tandem, the phase-resolved spectroscopy from JWST revealed longitudinally asymmetric cloud coverage on the dayside \citep{Coulombe2025}. Specifically, \planetname’s western hemisphere (morning side) was found to be shrouded in highly reflective clouds (local albedo $A \sim 0.8$), whereas its eastern hemisphere (evening side) is markedly darker ($A \sim 0.4$) \citep{Coulombe2025}. This remarkable cloud asymmetry, the western dayside being brighter, is exactly what one would expect if silicate clouds condense in cooler regions that have recently crossed from night to day. 
This asymmetry aligns with theoretical expectations that condensates preferentially form on the western terminator, where material advected from the nightside encounters cooler conditions at sunrise, favoring silicate cloud formation \citep{Mukherjee2025, Fu2025}.
\citet{Coulombe2025} propose that strong eastward winds transport heat toward the east side, keeping the western dayside slightly cooler and allowing condensate clouds to form there. The overall dayside albedo was measured around 0.5, and the thermal phase curve was found to be symmetric about the substellar point \citep{Coulombe2025}. The dayside brightness temperature (integrated over NIRISS bandpass) is $\sim2260$ K, while the nightside is extremely cold ($<1330$ K, 3$\sigma$ upper limit),  reaffirming very inefficient day-night energy redistribution and short radiative cooling times \citep{Coulombe2025}. These JWST results paint a picture of \planetname{} as a world with inhomogeneous, high-temperature clouds and an atmosphere that is both highly enriched and globally heterogeneous. 

Additional support for a high-metallicity, aerosol-laden atmosphere comes from ground-based high-resolution spectroscopy. \citet{Reyes2025} observed multiple transits of \planetname{} with VLT/ESPRESSO (covering 0.4-0.78 $\mu$m at high spectral resolution) in an attempt to detect atomic or molecular absorption lines through the planet’s terminator, however no significant atmospheric features were detected in the ESPRESSO data. The lack of observed optical absorption features is attributed to high-altitude clouds/hazes and a high mean molecular weight that together suppress spectral lines in the transmission spectrum \citep{Reyes2025}. Interestingly, the ESPRESSO data also showed no signs of significant atmospheric escape (e.g. no fast outflow of hydrogen), suggesting that \planetname’s atmosphere, while likely eroded over time, is presently stable,  an outcome that high metallicity and cloud cover may help to enable \citep{Reyes2025}. 

In this work, we present new JWST observations of \planetname{} that leverage the thermal infrared to probe the planet’s atmospheric composition and global circulation. Using JWST’s NIRSpec G395H instrument (covering $\sim$2.8-5.2 $\mu$m at $R\sim2700$), we obtained a spectroscopic phase curve of \planetname. These longer infrared wavelengths provide a more transparent window into the deeper atmospheric layers, because cloud opacity and stellar reflection both decrease toward the mid-IR. The planet’s own heat dominates the 3-5 $\mu$m flux, allowing us to observe what the reflective clouds had concealed at shorter wavelengths. The G395H phase curve unveils clear spectral signatures of carbon-bearing species across the planet’s orbit. In particular, we detect strong absorption features from CO and CO$_2$, confirming their global presence in \planetname’s hot atmosphere. Notably, CO (with a fundamental band at \sim4.7 $\mu$m) is seen even on the nightside, indicating that the nightside temperature remains high enough for CO to remain the dominant carbon carrier, as opposed to CH$_4$. We also find evidence of H$_2$O vapor, which appears most prominently when viewing the warmer dayside hemisphere. This suggests that water vapor is abundant, but on the cooler nightside it may partly condense or become sequestered beneath cloud layers, making its spectral features less apparent. Perhaps most intriguingly, our spectrum shows a tentative absorption feature consistent with SO$_2$ around $\sim$4 $\mu$m, a known photochemical product in hot, irradiated atmospheres. While SO$_2$ was notably detected in hot Jupiter WASP-39 b \citep{Tsai2023}, it was recently detected in the nightside of WASP-17 b at a significance of 2.3$\sigma$ \citep{LustigYaeger2025}. While the tentative SO$_2$ detection in this work is marginal and will require confirmation, its possibility is exciting. A low-abundance detection of SO$_2$ in \planetname’s atmospheric composition would provide crucial bridges towards our understanding of these photochemical cycles across exoplanet regimes and better comprehend complex star-planet processes \citep{crossfield:2023so2}. 

Our JWST/NIRSpec phase-curve observations provide the most detailed look at \planetname’s molecular composition and its variation from day to night. Observing the thermal emission at multiple longitudes, we probe the pressure-depth of clouds and how chemical abundances change with temperature across the planet. Leveraging the more thermal nature of NIRSpec G395H observations compared to the optical/NIR ranges offered by SOSS, we are now able to see deep into \planetname’s atmosphere beyond optically opaque clouds. Using heat to reveal what clouds concealed, we unveil the rich carbon chemistry and asymmetric climate of a planet remarkably continuing to survive the hot Neptune desert.

%\section{Methods} 
%\label{sec:methods} 

\section{Data} 
\label{sec:data}

\subsection{JWST Observations}
\label{sec:data:jwst}

We conducted phase curve observations of \planetname{} using JWST's NIRSpec G395H Bright Object Time Series (BOTS) mode, providing nearly continuous wavelength coverage from $2.8$--$5.2$~$\mu$m at a maximum spectral resolution of $R\sim2700$ \citep{Birkmann2022}.  Our program (JWST GO-3231) is designed to capture a full orbital phase curve, encompassing a transit, two secondary eclipses, and intermediate orbital phases required to probe longitudinal variations in the planet's atmosphere. 

% This observing strategy allows us to measure transmission, dayside emission, and longitudinally resolved thermal spectra within a single dataset. 

Each phase curve sequence spans a complete $\sim$19-hour orbital period, with the exposures bracketed by eclipses to provide robust constraints on instrument systematics and stellar variability. Observations are carried out using the NIRSpec BOTS aperture ($1.6''\times1.6''$), with an offset target acquisition is performed using a nearby star separated by $24''$. The host star and target, LTT 9779, (J=12.8) meets NIRSpec acquisition guidelines, with only a single predicted saturated pixel during wide-angle acquisition. Archival imaging and Gaia constraints confirm the absence of contaminating companions within $2''$ down to $\Delta K_s < 7$~mag. 

For this program, the observations are scheduled with a one-hour phase-constrained window around the predicted ephemerides, followed by continuous coverage of the entire orbit. This comprehensive observational approach yields access to the planet’s dayside, nightside, terminator, and intermediate longitudes, enabling unprecedented constraints on its atmospheric composition, cloud distribution, and global thermal structure.

\subsection{Data Reduction}
\label{sec:data:reduction}

We adopt two complementary reduction strategies to assess the reliability of our results. In the first approach (v1), the full calibration and extraction process is carried out within \eureka \citep{Bell2022, Ashtari2025}, from detector-level data through to the final spectroscopic light curves. In the second approach (v2), the raw data are initially processed with the \exotedrf pipeline \citep{Radica2024_JOSS,Feinstein2023,Radica2023}, which produces calibrated products through the equivalent of \eureka's Stage 4. These are subsequently ingested into \eureka for light-curve construction and spectral generation in Stages~5–6. Transit and eclipse modeling is then performed with the \texttt{batman} framework \citep{Kreidberg2015}. This dual-path analysis provides a consistency check between two distinct reduction chains, allowing us to quantify the sensitivity of our results to methodological choices.

\subsubsection{\eureka}
\label{sec:data:Eureka}

The primary \eureka data reduction is performed using the \eureka\ v1.1.2 JWST pipeline \citep{Bell2022, Ashtari2025}. All calibration and extraction steps are carried out internally with \eureka's modular stages (Stage 1 through Stage 6). As in previous applications of \eureka, we employ a prototype optimization framework \citep{Ashtari2025} to tune key parameters governing ramp fitting, background subtraction, spectral extraction, and outlier rejection.  

Stage 2 is executed using fixed default values from the Eureka! control file (\texttt{.ecf}), while Stages 1, 3, and 4 undergo parametric optimization. For each parameter, the optimizer iteratively sweeps through a specified range or set of values, calculating a fitness score defined by the median absolute deviation (MAD) of the white light curves. The parameter choice minimizing the MAD is selected before proceeding to the next optimization step. This sequential strategy yields light curves with reduced scatter and robust rejection of outliers, as illustrated in \autoref{fig:optimization}.  

Once an optimized set of parameters is identified, the full dataset is reprocessed from Stage 1, ensuring a consistent reduction across the entire phase curve sequence. The resulting time-series products provide spectroscopic light curves spanning all orbital phases, which are subsequently used to construct the phase-resolved emission spectra of \planetname.

\subsubsection{\exotedrf }
\label{sec:data:ExoTEDRF}

Our secondary reduction with \exotedrf\ is performed using version 2.1.0 and follows the methodology outlined in \citet{Ahrer_2025} and the official \exotedrf\ tutorial\footnote{\href{https://exotedrf.readthedocs.io/en/latest/content/notebooks/tutorial_nirspec-g395h.html}{ExoTEDRF tutorial}}. Stage~1 calibration begins with the standard detector-level step. We then apply a group-level $1/f$ noise correction utilizing the median value of each column, with separate treatments for even and odd detector rows. Pixels within 16 pixels (8 above and 8 below) of the target spectral trace, as well as all pixels flagged in a previous reduction, are masked during this step. A time-domain cosmic-ray rejection is performed with a $10\sigma$ clipping threshold.

In Stage~2, we repeat the $1/f$ noise and background subtraction at the integration level, keeping the same parameters as before. We interpolate any remaining bad pixels with a $10\sigma$ threshold in both spatial and temporal domains. We also run \exotedrf's principal component analysis (PCA) step, which reveals some outlier integrations; however, we do not subtract components directly and instead clip outlier integrations in the light-curve fitting stage. The spectral traces are then located using the edge-trigger algorithm of \citet{Radica_2022}, and the 1D spectra are extracted with a box aperture of width 8 pixels.

We also process the \planetname{} phase curve dataset using a hybrid reduction chain that combines \exotedrf \citep{Radica2024_JOSS} with \eureka\ v1.1.2 \citep{Bell2022, Ashtari2025}. In this approach, detector-level calibration through the equivalent of \eureka's Stage 4 is performed within \exotedrf, handling ramp fitting, background subtraction, flat-fielding, and wavelength calibration. The resulting products (uncalibrated 2D lightcurves) are then ingested into \eureka\ beginning at Stage 4.  

Spectroscopic light curves are constructed and optimized within \eureka\ using the same prototype parameter optimizer described in \cite{Ashtari2025}, but restricted to Stage~4 extraction parameters that impact the construction of 1D spectra and subsequent time-series light curves. As with the fully \eureka-based reduction, the optimizer evaluates parameter sets by sweeping through pre-defined ranges and selecting values that minimize the median absolute deviation (MAD) of the resulting light curves. This ensures consistent outlier rejection and minimized scatter in the final spectroscopic time series.  

\subsection{White Light Curve Fitting} 
\label{sec:wlc_fit:whiteLC} 

Once optimal extraction parameters are identified, \eureka Stages 5–6 are run on the both v1 and v2 datasets, producing light curves and spectra across six orbital phases. This hybrid pipeline provides an independent reduction path, while allowing direct comparison with the fully \eureka-based analysis and helping to isolate any sensitivity of the results to earlier data reduction choices.

The planet-to-star flux ($F_p/F_s$), radius ($R_p/R_s$), orbital period, mid-transit time ($t_0$), secondary eclipse time ($t_{sec}$), inclination ($i$), and scaled semi-major axis ($a/R_s$) are treated as free parameters and fitted using \eureka's MCMC-analysis tool (see \autoref{tab:fit_parameters}). Priors on the orbital parameters are informed using the values provided by \citet{Saha2025}, ensuring up-to-date and physically-consistent solutions. The orbital period is fixed to the best-fit \eureka\ value for the \exotedrf $+$ \eureka dataset, as allowing the period to vary reduced the time-correlated (red) noise below the theoretical Allan limit (see \autoref{fig:correlated_noise}). Eccentricity ($e$), argument of periastron ($\omega$), and  stellar radius ($R_p/R_s$) are fixed to \citet{Jenkins2020} values. Limb-darkening coefficients ($u_1$, $u_2$) were also fitted uniformly from 0 to 1 using a quadrature limb-darkening scheme, however spectroscopic fitting employs a fixed \texttt{exotic-ld} mps2 stellar limb-darkening model \citep{Grant2024}.  Systematics were modeled with free ramp parameters ($r_0$, $r_1$) describing an exponential decay, and the phase curve signal was captured with sinusoidal terms ($A_{\cos,1}$, $A_{\sin,1}$) representing modulation near secondary eclipse and quadrature. Together, this framework allowed robust estimation of the astrophysical signal while controlling for time-dependent detector effects.

\begin{figure}[]
    \centering
    \includegraphics[width=1.0\linewidth, trim={0cm 0cm 0cm 0cm}, clip]{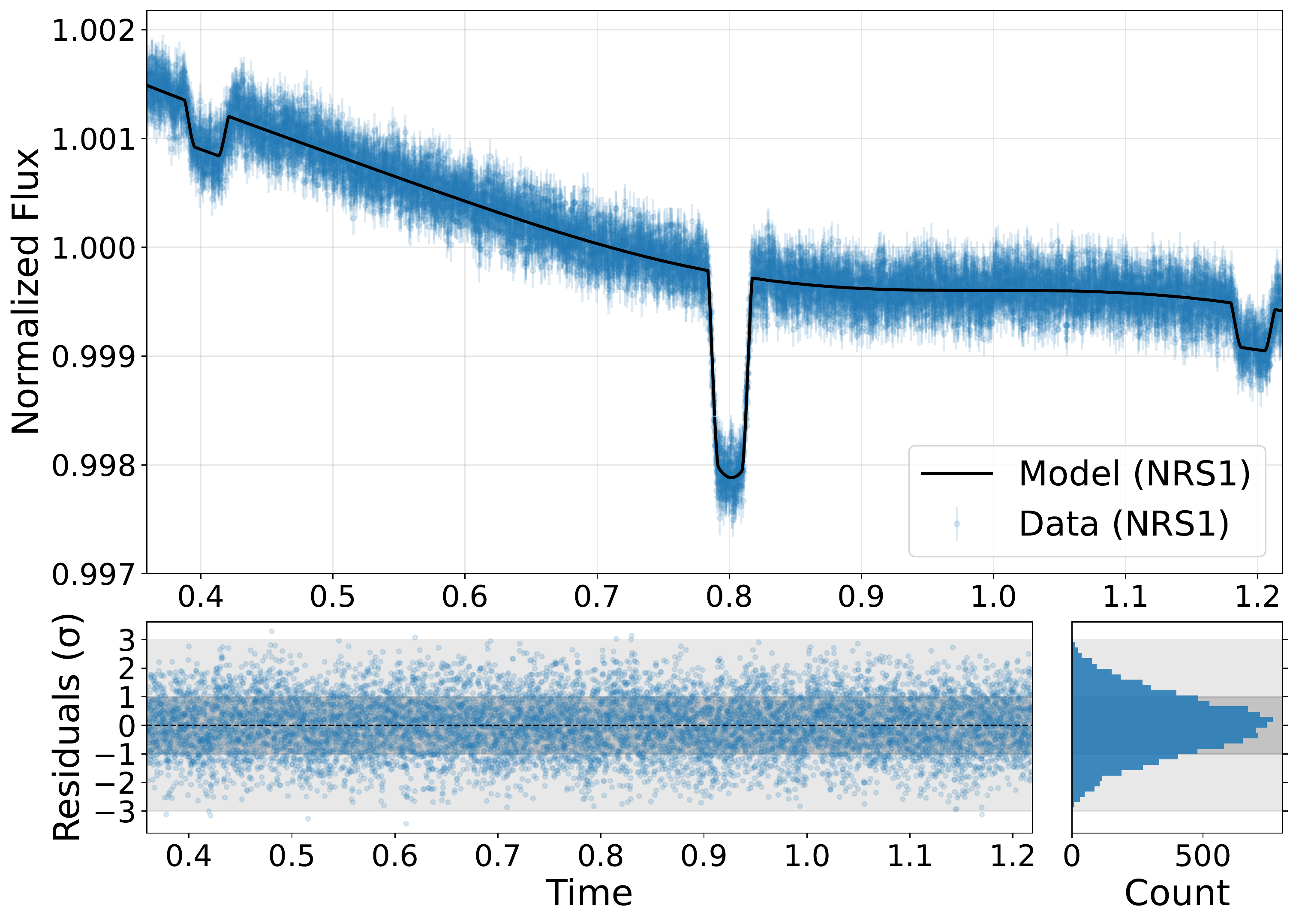}
    \vspace{0.5em}
    \includegraphics[width=1.0\linewidth, trim={0cm 0cm 0cm 0cm}, clip]{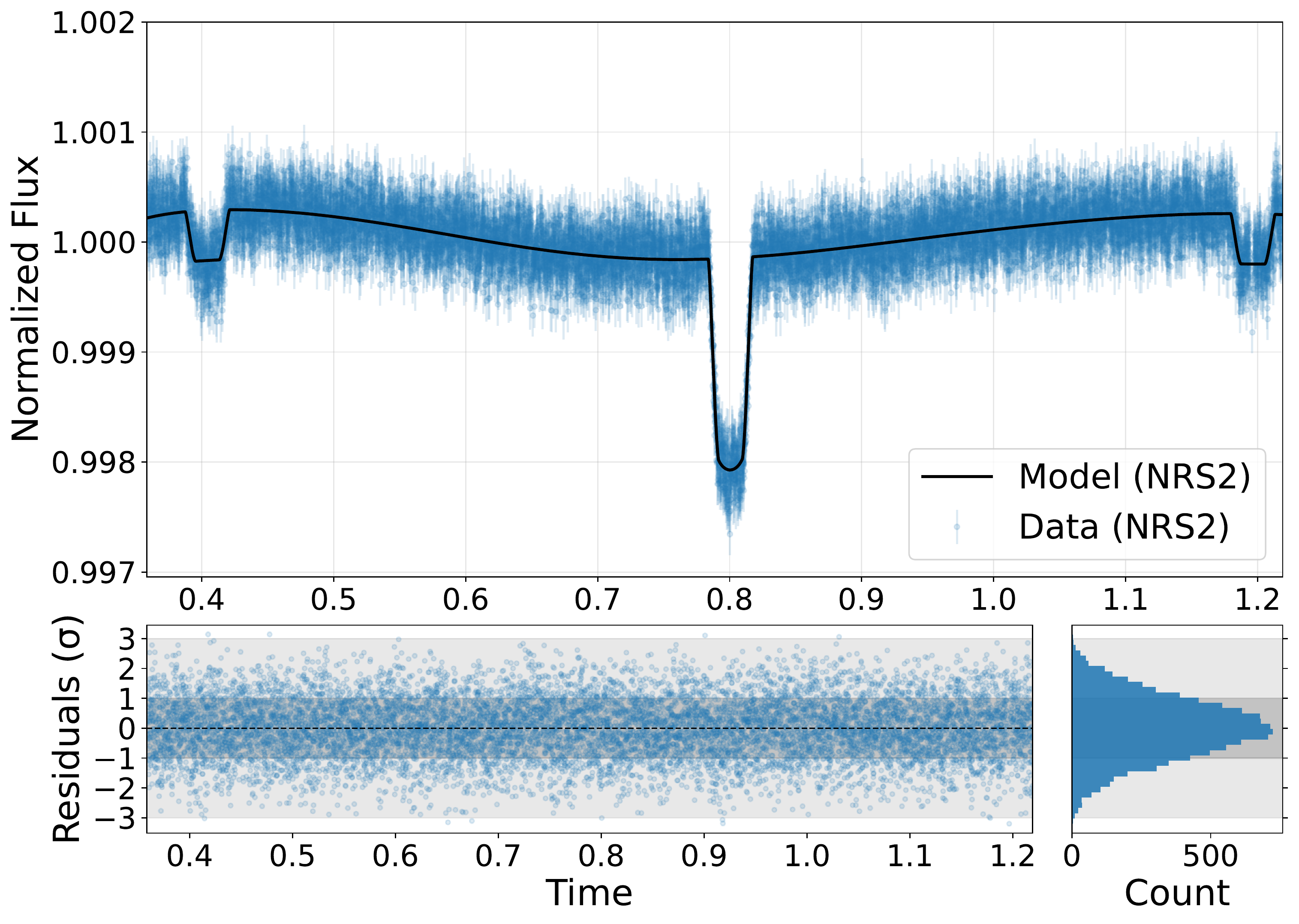}
    \caption{Broadband light curves, phase curve models, and residuals for the \eureka (v1) data reductions. NRS1 and NRS 2 detector data is shown on top and bottom. While NRS1 data demonstrated a significant ramp for both v1 and v2 datasets, systematic models were able to compensate and fit lightcurve data adequately.}
    \label{fig:LC_Fits}
\end{figure}

Broadband (white) light curves are well fit for both reductions. The \eureka-only dataset (v1) produces consistent NRS1 and NRS2 white light curves (\autoref{fig:LC_Fits}), and the same fitting procedure applied to the \exotedrf~+~\eureka\ dataset (v2) yields nearly identical results (\autoref{tab:fit_parameters}). Transit depths, eclipse depths and orbital parameters agree to within $1\sigma$.  

Differences are limited to the limb darkening coefficients, where datasets v1 and v2 favor slightly offset values. A dedicated LD model comparison (\autoref{fig:ld_comparison}) shows these differences have no impact on the fits, and both datasets are analyzed using fixed LD coefficients from the \texttt{exotic-ld} mps2 stellar model.

\subsection{Spectroscopic Light Curve Fitting} 
\label{sec:data:specLC} 

Using the best-fit system parameters from \autoref{tab:fit_parameters}, spectroscopic light curves are generated for both reductions across NRS1 and NRS2. Channelization is performed with 20\,nm bins, yielding 42 spectroscopic channels from 2.88-3.72\,$\mu$m (NRS1) and 66 channels from 3.83-5.13\,$\mu$m (NRS2) with a resultant resolution of $R$\sim$100$. 

In the spectroscopic fits, $F_p/F_s$ and $R_p/R_s$ vary with uniform priors, while the orbital period, $t_0$, $t_{\rm sec}$, inclination, and $a/R_s$ remain fixed to the white light solutions and system parameters. Eccentricity, argument of periastron, and $R_s$ are likewise fixed. The baseline terms $c_0$, $c_1$, and a scatter multiplier vary freely. The scatter multiplier ($scatter\_mult$) serves as a white-noise scaling factor, multiplying the Stage 3 uncertainties to account for under or over-estimated noise levels, and is the recommended method for capturing residual variance in \eureka\ fits. Time-correlated systematics are modeled with an exponential ramp ($r_0$, $r_1$), and sinusoidal modulation terms ($A_{\cos,1}$, $A_{\sin,1}$) are included to capture orbital phase brightness variations. This setup balances fixed orbital geometry with flexibility in the photometric baseline and phase curve structure, yielding stable spectroscopic fits across all channels.

Limb-darkening priors are set using the \texttt{recenter\_ld\_prior} feature in \eureka, with values anchored to the \texttt{exotic-ld} stellar grids. A comparison of fitted limb-darkening coefficients against the MPS2 model across all spectroscopic channels is shown in \autoref{fig:ld_comparison}, confirming that MPS2 provides the most consistent description of the stellar intensity profile. Both v1 and v2 analyses therefore adopt fixed MPS2 coefficients for all spectroscopic fits.  

Phase-resolved spectroscopic light curves are extracted at six orbital phases: 0$^{\circ}$ (transit), 60$^{\circ}$, 120$^{\circ}$, 180$^{\circ}$ (secondary eclipse), 240$^{\circ}$, and 300$^{\circ}$. Comparisons of the resulting planet-to-star flux ratios for v1 and v2 are presented in \autoref{fig:reduction comparison}. Across all phases and channels, the two reductions agree to within $\sim$1$\sigma$.

The agreement is further reflected in the limb-darkening fits, where the MPS2-based coefficients track the spectroscopic behavior of both v1 and v2 datasets closely (see \autoref{fig:reduction comparison}). With consistent results across all orbital phases, the spectroscopic reductions provide a reliable basis for the subsequent retrieval analysis of\planetname’s global atmospheric properties.

\section{Modeling} 
\label{sec:modeling}

\begin{figure*}[t]
    \centering
    \includegraphics[width=1.0\linewidth, trim={0cm 0cm 0cm 0cm}, clip]{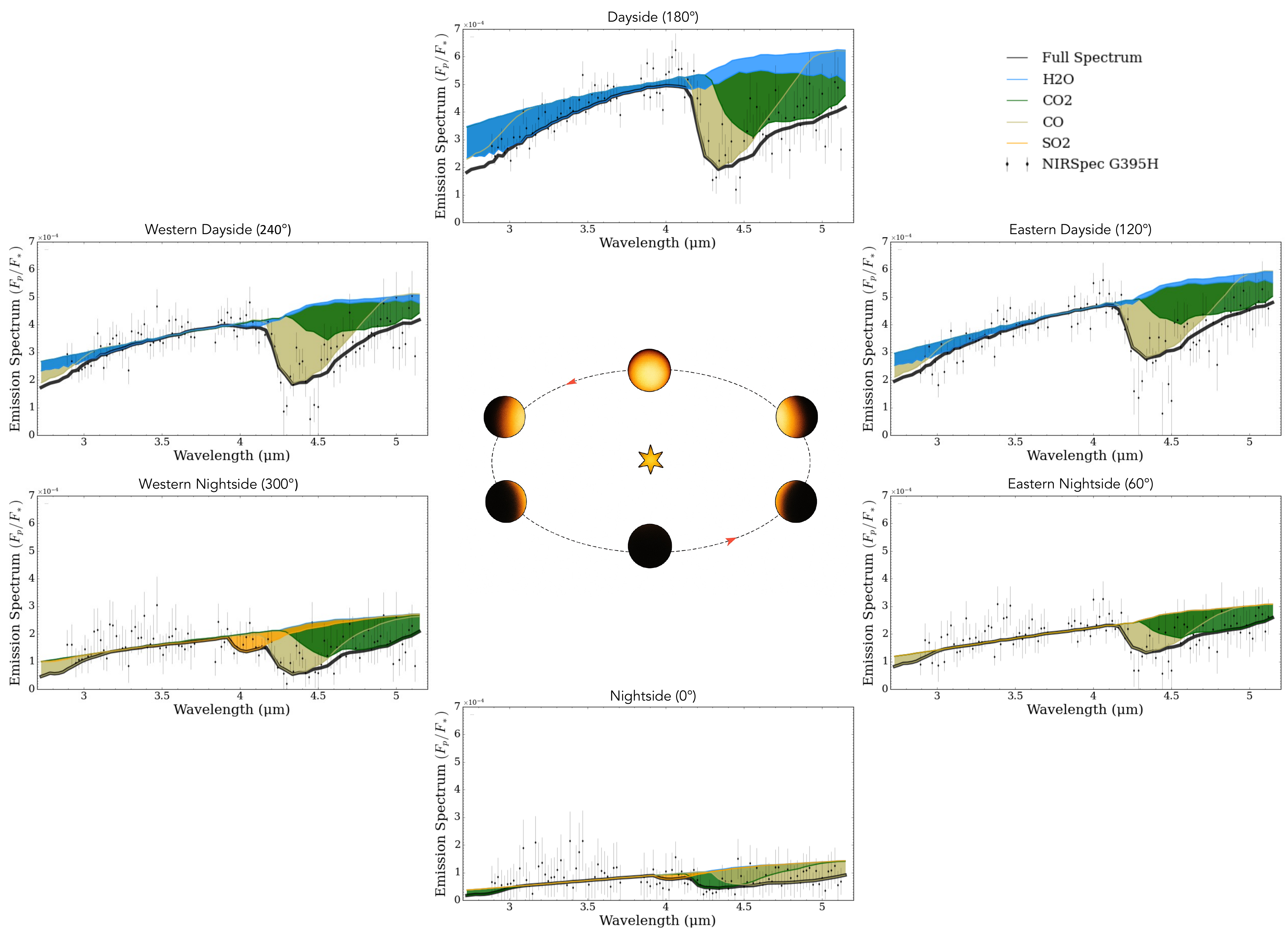}
    \caption{Emission spectra and best-fit free retrieved models for all six phases. The spectral contributions of CO, CO$_2$, H$_2$O and SO$_2$ abundances are shown in green, gold, blue and orange for each phase. The best-fit, full atmospheric models for the phase-resolved emission spectra are shown in black.}
    \label{fig:Spectral_Decomposition}
\end{figure*}

\begin{figure*}[t]
    \centering
    \includegraphics[width=1.0\linewidth, trim={0cm 0cm 0cm 0cm}, clip]{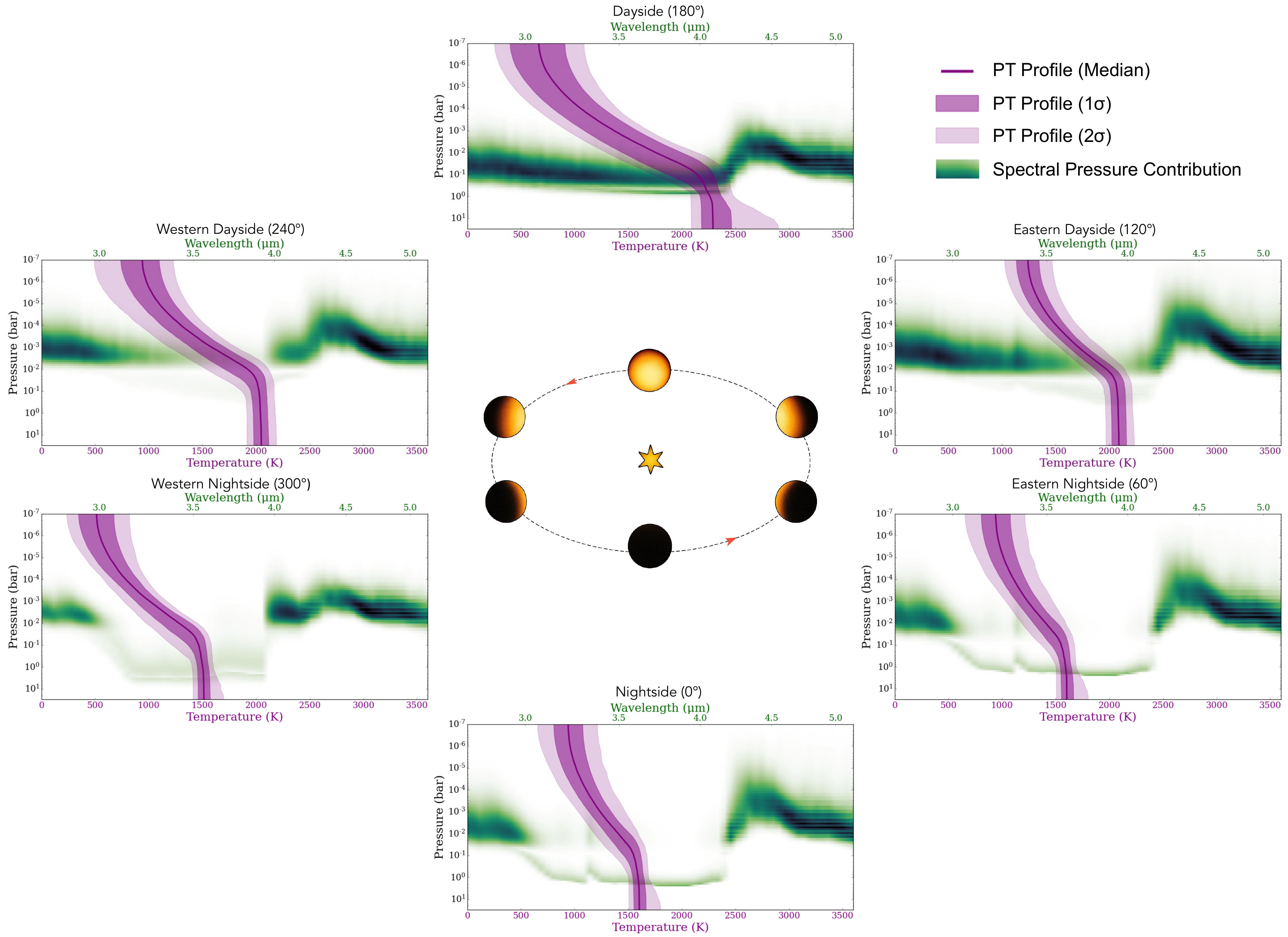}
    \caption{Retrieved atmospheric P-T profiles and spectral pressure contributions for all six phases. Pressure temperature profiles are shown in purple, corresponding to the y-axes and bottom x-axes of each subplot. Spectral contributions are shown in green, corresponding to the y-axes and top x-axes of each subplot. Sharing a pressure axes, we are able to analyze how deep into the atmosphere we are probing at each phase. Notably, the intermediary phases present a lower-pressure depth during emission spectroscopy, potentially indicating high-altitude clouds.}
    \label{fig:Pressure_Contribution}
\end{figure*}

\begin{figure*}[t]
    \centering
    \includegraphics[width=1.0\linewidth, trim={0cm 0cm 0cm 0cm}, clip]{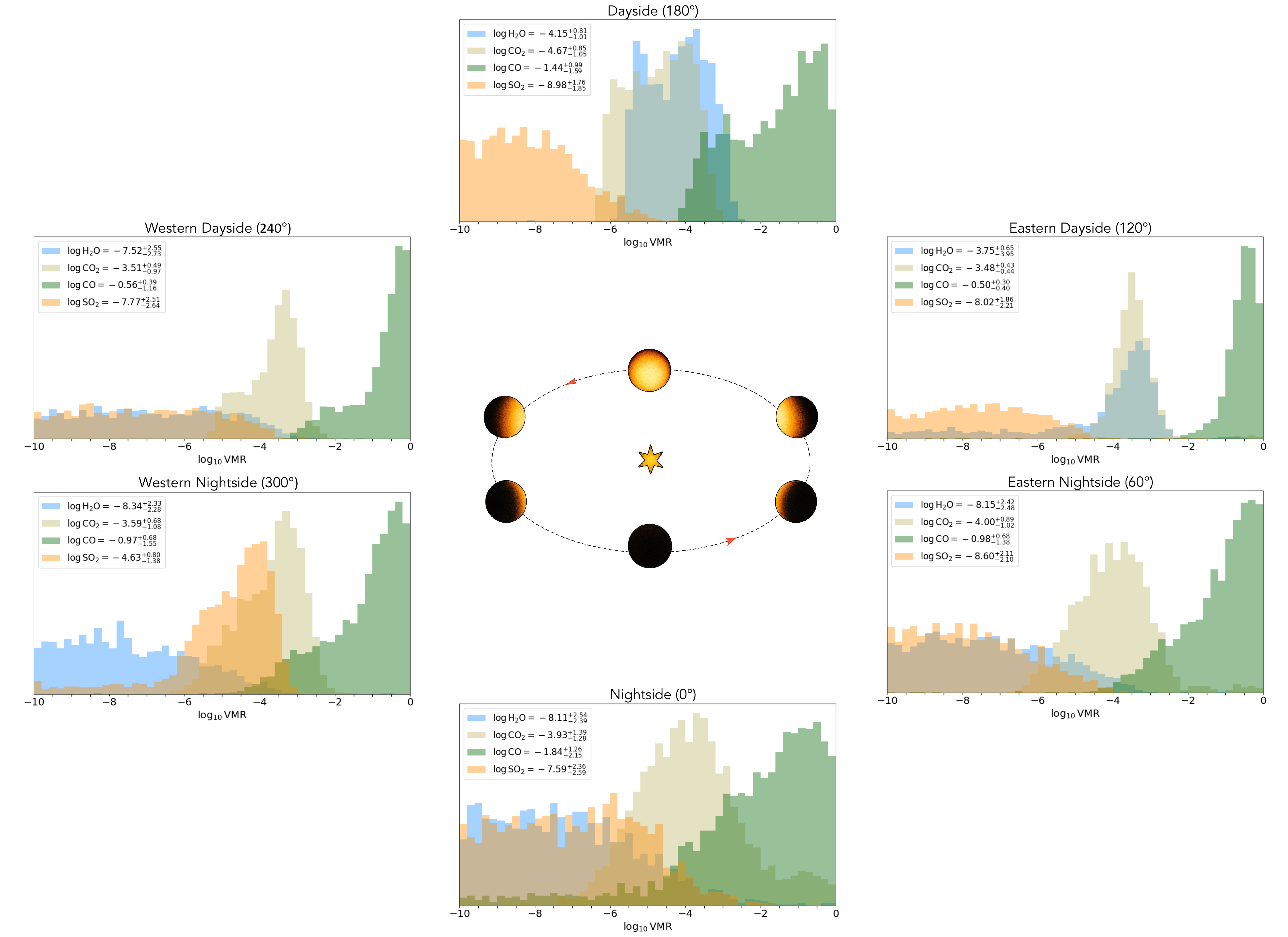}
    \caption{Retrieved CO, CO$_2$, H$_2$O and SO$_2$ abundances for all six orbital phases. Carbon-bearing species remain prominent through-out all phases. H$_2$O is detected in the Eastern Dayside and Dayside, but apparently absent elsewhere. Given a lack of evidence for thermal dissociation, the H$_2$O is still likely present in the atmosphere, but possibly concealed by higher altitude clouds as explained in \autoref{fig:Pressure_Contribution}. SO$_2$ is tentatively present in the Western Nightside of the planet, hinting at possible photochemistry.} 
    \label{fig:Abundances}
\end{figure*}

\subsection{Atmospheric Retrievals}
\label{sec:modeling:retrievals}

\subsubsection{Retrieval Setup}

We use the publicly available \texttt{POSEIDON} software program to perform free retrievals of \planetname{}'s atmosphere \citep{MacDonald2017, MacDonald2023}. We perform the retrievals at a spectral resolution of $R \sim 20{,}000$, employing the \texttt{MultiNest} nested-sampling algorithm with 800 live points to explore the parameter space efficiently. 

To model the pressure-temperature (PT) profile of the planet's atmosphere, we implement the parameterized three-part profile of \citet{Madhu2009}. This three-part PT profile is implemented with uniform priors on all free parameters: $a_1$, $a_2$, $\log P_1$, $\log P_2$, $\log P_3$, and the reference temperature $T_\mathrm{ref}$. The thermal gradient parameters $a_1$ and $a_2$ vary between 0.02 and 2.0, while the pressure breakpoints span wide ranges to allow flexibility in locating the photosphere: $\log P_1, \log P_2 \in [-7,2]$ and $\log P_3 \in [-2,2]$, with pressures expressed in bar. The reference temperature prior extends from 200 to 4000~K, accommodating both cool nightside and hot dayside solutions. This setup ensures the retrieval explores a broad but physically motivated parameter space, allowing the data to constrain the temperature structure without overly restrictive assumptions. 

We also evaluate a Guillot profile \citep{Guillot2010}; however, these solutions produce temperature gradients that exceed the adiabatic limit expected for \planetname\ \citep{Parmentier2014}. The maximum physically allowed PT gradient is set by the adiabatic relation \citep{Parmentier2014}. For an ideal gas, the Poisson relation
\begin{equation}
    T\,P^{\frac{1-\gamma}{\gamma}} = \mathrm{constant},
\end{equation}
links the temperature $T$, pressure $P$, and the ratio of specific heats $\gamma = c_p/c_v$ (with $c_p$ and $c_v$ the specific heat capacities at constant pressure and volume, respectively). Differentiating in base-10 logarithms gives
\begin{equation}
    \frac{d\log_{10} T}{d\log_{10} P} = \frac{\gamma-1}{\gamma}.
\end{equation}
This form yields a dimensionless adiabatic gradient. Rearranging the equation with respect to $\log_{10} P$ produces
\begin{equation}
    \frac{dT}{d\log_{10} P} = \ln(10)\,\frac{\gamma-1}{\gamma}\,T,
\end{equation}
where $\ln(10)$ converts between natural and base-10 logarithms.  

For a hydrogen-dominated atmosphere (mean molecular weight $\mu \approx 2.3$) with $\gamma \simeq 1.3$ at $T=2000$~K, this expression gives
\begin{equation}
    \frac{dT}{d\log_{10} P} \;\approx\; 1.1\times10^{3}~\mathrm{K\ dex^{-1}}.
\end{equation}
This value provides the benchmark we adopt when assessing whether retrieved PT profiles are physically consistent with convective stability.

To analyze the chemical composition of \planetname, we retrieve the abundances of \ce{H2O} \citep{Polyansky2018}, \ce{CH4} \citep{Yurchenko2024}, \ce{CO2} \citep{Yurchenko2020}, \ce{CO} \citep{Li2015}, and \ce{SO2} \citep{Underwood2016}.  These species represent the canonical set of molecules expected in hot Jupiter atmospheres (e.g., WASP-39b; \citealt{Alderson2023, Rustamkulov2023, Tsai2023}), as they are predicted by photochemical grid models and have been robustly detected in comparable planets. We also evaluate a more extensive subset of molecules and atoms to test for high-temperature dissociation. No evidence for \ce{H-}, \ce{OH}, or \ce{OH+} is found throughout the orbit of the planet. 

Assessing each orbital phase for cloudy versus clear conditions, all six spectra favor cloud-free models. This outcome is unsurprising, as the infrared sensitivity of NIRSpec/G395H, compared to previous NIRISS/SOSS observations of \planetname, is less suited to constraining cloud features.

\subsubsection{Retrieval Results}

Our atmospheric retrieval of \planetname’s G395H phase-resolved emission spectra reveals a high-metallicity, carbon-rich composition with pronounced longitudinal variations in cloud coverage and chemistry, specifically identifying four key molecular species across orbital phases: CO$_2$, CO, H$_2$O and SO$_2$.

CO$_2$ is robustly detected and well-constrained at all longitudes. The phase-resolved spectra show a prominent CO$_2$ absorption band near 4.3~$\mu$m at every phase, confirming a substantial CO$_2$ presence on both dayside and nightside. The combination of abundant CO$_2$ and appreciable (if prior-limited) CO indicates that \planetname’s atmosphere is globally enriched in heavy elements, i.e. a high-metallicity atmosphere. A similar inference was made for JWST’s ERS target WASP-39b, which exhibited strong CO$_2$ absorption and implied a super-solar heavy-element content \citep{Alderson2023}.

Retrieved CO abundance is only weakly constrained at most orbital phases, with posterior distributions abutting the upper prior bound. Thus, we interpret CO abundances in terms of their median values, treating them effectively as upper limits. This behavior mirrors other JWST retrieval studies (e.g., WASP-39b and WASP-18b), where CO often remains unconstrained and prior-limited \citep{Alderson2023, Coulombe2023}. 

In our retrieval, with an upper limit mixing ratio of $<10^{-7.6}$, we find no evidence for significant CH$_4$ or other reduced carbon species.%, suggesting that most atmospheric carbon resides in CO$_2$ (and CO), consistent with expectations for a $\sim$2000K regime with near-solar C/O ratio \citep{Visscher2006, Visscher2010}.
Our identification of CO$_2$, CO, and H$_2$O (and not CH$_4$) on \planetname{}'s dayside appears to be consistent with high-temperature atmospheres at super-Solar metallicities \citep{Moses2013b}. By using the median abundances of these species, we directly measure a C/O$=0.997$, consistent with predictions from \citet{Moses2013b}. Independent equilibrium chemistry modeling of \planetname{}'s dayside emission spectrum shows [Fe/H]~>~$500\times$~Solar and C/O~$\sim0.93$ \citep{Brande2025}.
 
\autoref{fig:Spectral_Decomposition} presents the measured emission spectra at representative orbital phases, decomposed into the contributions of major molecular absorbers. CO$_2$ dominates the spectral modulation at 4–5~$\mu$m for all phases, while CO’s spectral influence is comparatively minor and degenerate with the blackbody continuum, explaining why CO abundances are only weakly constrained. The persistent depth of the CO$_2$ feature across phases supports a globally high carbon, even on the planet’s nightside. 
 
Water vapor is detected on \planetname’s hot dayside, but its spectral signatures fade and essentially vanish toward the cooler nightside. In the high-S/N dayside spectrum, we observe broad H$_2$O absorption features in the 3.2–3.8$\mu$m range and contributing to the continuum depression shortward of 3.5~$\mu$m (\autoref{fig:Spectral_Decomposition}, dayside spectrum). By contrast, spectra from the nightside (phase $\sim$0.0/1.0) show a nearly gray, blackbody-like shape with no significant H$_2$O absorption trough at these wavelengths. Our retrieval quantitatively confirms this dichotomy: the retrieved H$_2$O mixing ratio drops by ${>1}~\mathrm{dex}$ from the dayside to the coldest nightside longitudes. We attribute this strong longitudinal H$_2$O asymmetry to cloud formation and opacity effects rather than to thermal dissociation of water. 

The planet’s nightside is sufficiently cool (effective $T{\rm eff}\lesssim1300$K at 3$\sigma$) for silicate and oxide clouds to condense \citep{Visscher2006, Visscher2010}, which would obscure gaseous H$_2$O signatures by forming a high-altitude cloud deck. \planetname’s nightside is inferred to be cloud-dominated in optical/NIR observations \citep{Coulombe2025}, and our emission spectra provide indirect evidence of clouds blanketing the cooler hemispheres. 

Further insight into the cloud structure of the planet is provided by the retrieved atmospheric P–T profiles and spectral pressure contributions in \autoref{fig:Pressure_Contribution}. P-T profiles are shown in purple, corresponding to the y-axes and bottom x-axes of each subplot. Spectral contributions are shown
in green, corresponding to the y-axes and top x-axes of each subplot. Sharing a common pressure axis allows us to assess how deep into the atmosphere is probed at each phase. Notably, the intermediary phases present lower-pressure depths during emission spectroscopy, consistent with the presence of high-altitude clouds. These same phases were also identified in the NIRISS/SOSS phase curve of \citet{Coulombe2025}, where reflected-light spectroscopy revealed enhanced cloud scatter on the western hemisphere. Reinforcing the SOSS team's analysis, our G395H retrievals support the interpretation that high-altitude condensates dominate at intermediate longitudes, likely culprit to both muting H$_2$O signatures in emission and the enhanced reflectivity inferred from the optical/NIR phase curve.

Crucially, we detect no spectral signatures of water dissociation products—namely H$^{-}$ continuum opacity or emission from OH/OH$^+$ radicals—at any phase. If water were being thermally destroyed on the dayside, one would expect a strong optical/near-IR H$^{-}$ slope or hydroxyl emission at these temperatures ($\sim2000$K). The absence of such features in \planetname’s spectra indicates that H$_2$O is not significantly dissociated; instead, its apparent disappearance is consistent with condensation into clouds on the cooler nightside. This interpretation aligns with findings on other tidally locked hot planets: for example, JWST phase-curve spectra of the hot Jupiter WASP-43b require thick nightside clouds to explain the suppressed molecular features and low flux from the nightside \citep{Bell2024}. In \planetname’s case, water vapor likely remains present in gaseous form on the hotter western dayside (where we see pronounced absorption features), but is obscured by higher altitude clouds. \autoref{fig:Pressure_Contribution} illustrates the retrieved thermal profiles and emission contribution functions at the day and night limbs. The dayside temperature–pressure ($T$–$P$) profile is non-inverted and hot ($T\approx2200$K at 0.1bar), allowing gaseous H$_2$O to exist throughout the near-IR photosphere. On the nightside, the $T$–$P$ profile is far cooler ($T<1300$K at 0.1bar) and features a pronounced temperature gradient; correspondingly, the nightside contribution function peaks at higher pressures ($\sim$0.3–1bar) where clouds are expected to form. These results underscore that cloud opacity, not chemical depletion, is the primary reason for the lack of H$_2$O features in the cooler phases.

In the Western Nightside ($\sim$300$^\circ$), corresponding to the western terminator/nightside region, we retrieve a tentative spectroscopic feature attributable to sulfur dioxide (SO$_2$). This SO$_2$ signal manifests as an extra absorption bump near $\sim$4$\mu$m in the phase-300$^\circ$ spectrum (\autoref{fig:Spectral_Decomposition}, green curve), which is not required by models lacking SO$_2$. When included in the retrieval, the SO$_2$ volume mixing ratio at this phase is constrained to $\sim$10$^{-6}$–10$^{-5}$ (with large uncertainty), and the Bayes factor indicates a $\approx2.4\sigma$ preference for the SO$_2$ feature’s presence \citep{kipping:2025}. It is worth noting these detection significances are an upper limit for detection, as explained in \citet{kipping:2025}. SO$_2$ is not detected at other phases except for a weak $\approx1.1\sigma$ presence on the nightside (significances listed in \autoref{tab:significances}). This is consistent with the expectation that SO$_2$ would be produced on the cooler, cloudier limb where photochemistry can act on gas transported from the dayside \cite{Tsai2023}. If confirmed, this represents the first identification of SO$_2$ in the emission spectrum of a hot or ultra-hot Neptune. Notably, JWST has recently detected SO$_2$ in transmission spectra of several hot/warm gas giants. The prototypical case is WASP-39b, a $\sim$1100K hot Saturn where SO$_2$ was robustly identified at 4.05$\mu$m in the near-IR and at 7.7/8.6~$\mu$m in the mid-IR \citep{Powell2024,Rustamkulov2023,Tsai2023}. More unexpectedly, SO$_2$ (along with H$_2$O and silicate clouds) was also found in the cooler Neptune-mass planet WASP-107b (effective $T\sim740$K) via JWST/MIRI \citep{Dyrdek2024, Welbanks2024} and warm Neptune GJ~3470b (effective $T\sim 681$K) via JWST/NIRCam \citep{Beatty2024}. Those detections established SO$_2$ as a photochemically-produced species: ultraviolet irradiation of H$_2$S can generate SO$_2$ in planetary atmospheres. 

In \planetname’s atmosphere, the tentative SO$_2$ absorption at the western terminator likely traces a similar process. The planet’s host star is relatively bright in the ultraviolet, and the west limb (evening terminator) receives the stellar flux that has traveled through the dayside, enabling photolysis-driven chemistry on that limb. The retrieved SO$_2$ abundance (if real) suggests an atmosphere enriched in sulfur-bearing species and exposed to high-energy radiation, consistent with the planet’s high metallicity and intense irradiation \citep[cf.\ ][]{Crossfield2025}. We caution that the SO$_2$ detection in \planetname{} is marginal; additional spectra or repeated phase-curve observations will be needed to verify it. Nonetheless, its possible presence is intriguing in the broader context of exoplanet atmospheres. Like the SO$_2$ in WASP-39b and WASP-107b, it may point to active photochemistry and high atmospheric metallicity on \planetname{} \citep{crossfield:2023so2,Brande2025}.

\autoref{fig:Abundances} summarizes the retrieved molecular abundances for the key species as a function of orbital phase. CO$_2$ is consistently retrieved at a high mixing ratio ($\sim$10$^{-4}$–10$^{-3}$) across all longitudes, reinforcing the picture of a carbon-enriched atmosphere. CO posteriors (green regions in \autoref{fig:Abundances}) are prior-limited on the dayside and terminator, such that only upper limits (of order $\lesssim$10$^{-3}$ by volume) can be placed on CO. H$_2$O abundances (blue regions) drop precipitously from the dayside (where $\sim$10$^{-4}$ is retrieved) to the nightside ($<10^{-6}$ at 1$\sigma$ upper limit), quantitatively demonstrating the strong suppression of water vapor on the nightside due to cloud formation. Interestingly, the retrieved SO$_2$ abundance shows a significant spike at the 300$^\circ$ phase, reaching $\sim$10$^{-5}$, whereas at other longitudes only upper limits of order $10^{-7}$ or lower are obtained, a potential pattern consistent with a localized SO$_2$ feature confined to the cooler western nightside limb for highly-irradiated Jovian and Neptune-mass planets \citep{LustigYaeger2025}. This longitudinal distribution lends further credence to a photochemical origin for SO$_2$, as the terminator regions (especially the evening limb) are where such products are expected to accumulate \cite{Stamer2025}. Such a pattern is consistent with global circulation models predicting enhanced condensation on the western terminator, where atmospheric flow advects material from the hot dayside into cooler longitudes, creating favorable conditions for cloud formation \citep{Tsai2023, Mukherjee2025, Fu2025}.

A comparison between the free retrieval results and predictions from chemical equilibrium models for the planet's dayside is shown in \autoref{fig:VMR_comparison_free_chemeq}. The plotted pressure range (0 to $-3.5$ in $\log_{10}(P\,[\mathrm{bar}])$) corresponds closely to the emission contribution pressures probed in the dayside spectrum (top of \autoref{fig:Pressure_Contribution}). Across this region, equilibrium abundances for CO$_2$ align closely with the retrieved posterior distributions, and both retrieval methods produce consistently high CO$_2$ volume mixing ratios. For CO and H$_2$O, the equilibrium profiles fall near the upper limits of the free retrieval posteriors, lending support to the interpretation that the retrieved values are physically motivated upper bounds rather than poorly constrained fits. CH$_4$ and SO$_2$ abundances remain insignificant in both approaches. This overall agreement between the free and equilibrium frameworks reinforces the chemical plausibility of the retrieved atmospheric composition.

\subsection{Energy Budget}
\label{sec:modeling:budget}

\subsubsection{Phase-Resolved Effective Temperatures}

\begin{figure}[b]
  \centering
  \includegraphics[width=1\linewidth]{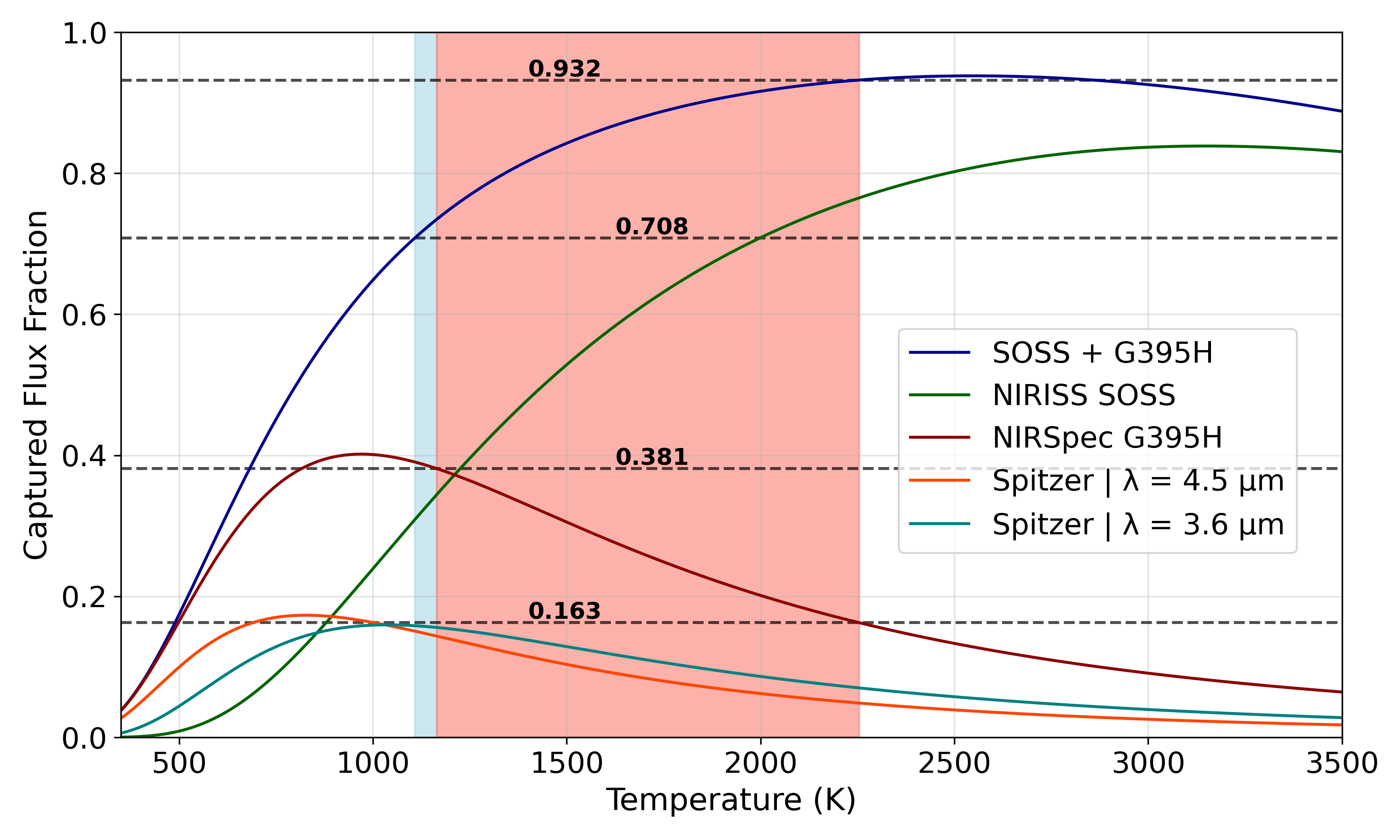}
  \caption{Captured flux fraction estimates of \planetname{} assuming blackbody emission. The red and blue boxes show the calculated effective temperature range of the planet using NIRSpec and NIRSpec+NIRISS, respectively. Using the same formulation as \citet{Splinter2025}, we estimate the captured flux fraction as the ratio of an instrument's band-integrated blackbody emission to the planet's bolometric emission. Black dashed lines indicate the captured flux fraction boundaries within the effective temperatures. NIRSpec/G395H captures $\sim$16–38\% of the bolometric flux while the combination of NIRSpec and NIRISS captures $\sim$71–93\% of the bolometric flux.}
  \label{fig:flux}
\end{figure}

Phase curves probe the longitudinal thermal structure of exoplanets \citep{Parmentier2018}. When phase curve measurements from NIRSpec/G395H and NIRISS/SOSS are combined, the wavelength range spans $\sim$71\ -- 93\% of the emitted flux (see \autoref{fig:flux}), providing rigorous constraints on the energy budget of \planetname{}. 

We use the same method as \citet{Splinter2025} to convert $F_p/F_s$ spectra to brightness temperatures. By rearranging the Planck function, the brightness temperatures
\begin{equation}
    T_{b} = \frac{hc}{k\lambda}  \left[\ln \left(\frac{2hc^2}{\lambda^5 B_{\lambda, \mathrm{planet}}} \right) +1 \right]^{-1} 
\end{equation}
can be related to $F_p/F_s$ by the thermal emission of the planet
\begin{equation}
    B_{\lambda, \mathrm{planet}} = \frac{F_p/F_s}{(R_p/R_\star)^2  B_{\lambda, \mathrm{star}}}.
\end{equation}

There are multiple approaches to convert brightness temperature spectra to effective temperatures including the error-weighted mean, power-weighted mean and linear interpolation \citep{Cowan2011_statistics, Schwartz2015}. We use the non-parametric Gaussian process regression of \citet{Pass2019} to estimate effective temperatures.

We adopt all stellar, planetary and orbital parameters and uncertainties from \citet{Jenkins2020}. Longitudinally resolved effective temperatures are computed from the NIRSpec phase curve alone and in combination with the median-fit thermal emission phase curve from NIRISS in \citet{Coulombe2025} (see \autoref{fig:teff_appendix} and \autoref{tab:system_parameters}). Additionally, assuming the planet radiates as a blackbody, we estimate the total captured flux fraction of the two bandpasses in \autoref{fig:flux}.

\begin{deluxetable*}{lcc}
\tablecaption{System parameters for \planetname. Stellar and planetary parameters are from \citet{Jenkins2020}, while energy budget values are from this work. We only list one irradiation temperature since it is independent of the NIRSpec and NIRISS phase curves.
\label{tab:system_parameters}}
\tablehead{\colhead{Parameter} & \colhead{Symbol} & \colhead{Value}}
\startdata
\multicolumn{3}{l}{\textit{Stellar}}\\
Stellar temp. & $T_\star$ & $5480 \pm 42~\mathrm{K}$ \\
Stellar radius & $R_\star$ & $0.949 \pm 0.006~R_\odot$ \\
\multicolumn{3}{l}{\textit{Planetary}}\\
Semi-major axis & $a$ & $0.01679^{+0.00014}_{-0.00013}~\mathrm{AU}$ \\
Radius & $R_p$ & $4.72 \pm 0.23~R_\oplus$ \\
\multicolumn{3}{l}{\textit{Energy Budget -- NIRSpec}}\\
Irradiation temp. & $T_\mathrm{irr}$ & $2783^{+35}_{-38}~\mathrm{K}$ \\
Dayside temp. & $T_\mathrm{day}$ & $2254 \pm 83~\mathrm{K}$ \\
Nightside temp. & $T_\mathrm{night}$ & $1165^{+49}_{-46}~\mathrm{K}$ \\
Recirc. efficiency & $\varepsilon$ & $0.17 \pm 0.03$ \\
Bond albedo & $A_B$ & $0.28 \pm 0.10$ \\
\multicolumn{3}{l}{\textit{Energy Budget -- NIRSpec+NIRISS}}\\
Dayside temp. & $T_\mathrm{day}$ & $2252^{+19}_{-17}~\mathrm{K}$ \\
Nightside temp. & $T_\mathrm{night}$ & $1109^{+23}_{-25}~\mathrm{K}$ \\
Recirc. efficiency & $\varepsilon$ & $0.14 \pm 0.01$ \\
Bond albedo & $A_B$ & $0.29 \pm 0.03$ \\
\enddata
\end{deluxetable*}

\subsubsection{Bond Albedo and Heat Recirculation Efficiency}

We estimate the Bond albedo $A_B$ and day-night heat recirculation efficiency $\epsilon$ of \planetname{} using the parametrization of the dayside temperature
\begin{equation}
    T_{\mathrm{day}} = T_0(1-A_b)^{1/4} \left(\frac{2}{3}  - \frac{5}{12}\epsilon \right)^{1/4}
\end{equation}
and nightside temperature
\begin{equation}
    T_{\mathrm{night}} = T_0(1-A_b)^{1/4} \left(\frac{\epsilon}{4} \right)^{1/4}
\end{equation}
from \citet{Cowan2011_statistics}. Here, $T_0 = T_\star\sqrt{R_\star/a}$ is the irradiation temperature, and $A_B$ and $\epsilon$ run from 0 to 1.

We estimate the best-fit values of $A_B$, and $\epsilon$ using a 1000 step Monte Carlo simulation where each iteration, the stellar, planetary and orbital parameters are perturbed by a Gaussian based on their respective uncertainties. From the NIRSpec phase curve, a Bond albedo of $A_B = 0.28 \pm 0.10$, and a heat recirculation efficiency of $\epsilon = 0.17 \pm 0.03$. Using the NIRSpec+NIRISS phase curve, we obtain more precise constraints: $A_B = 0.29 \pm 0.03$ and $\epsilon = 0.14 \pm 0.01$. Using this Monte Carlo simulation, we also infer an irradiation temperature of $T_0 = 2783^{+35}_{-38}~\mathrm{K}$.

\begin{figure}[tp]
  \centering
  \includegraphics[width=1.\linewidth]{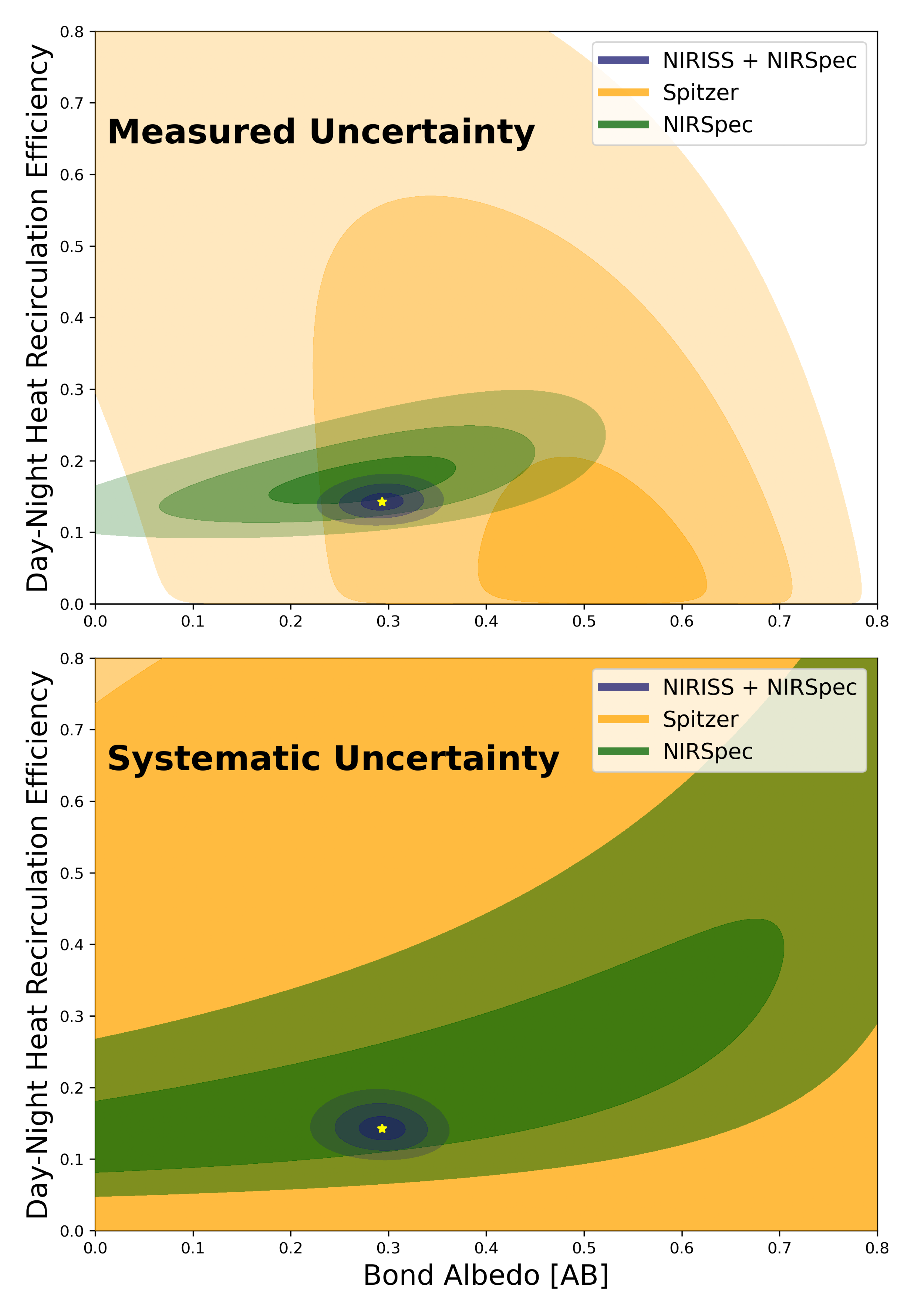}  % or .pdf, .jpg
  \caption{Bond albedo $(A_B)$ and heat recirculation efficiency ($\epsilon$) $\chi^2$ surfaces of \planetname{} calculated from dayside and nightside effective temperature estimates, and best-fit irradiation temperature $T_0 = 2783^{+35}_{-38}$ K. The $1\sigma$, $2\sigma$ and $3\sigma$ regions for each instrument are shown in a different shade. The yellow star represents the best-fit $A_B$ and $\epsilon$ as calculated from the 1000 iteration Monte-Carlo simulation with the NIRSpec+NIRISS effective temperatures. \textit{Top:} $\chi^2$ surfaces derived with standard errors. \textit{Bottom:} $\chi^2$ surfaces calculated with errors inflated by captured flux fractions. With errors inflated by the captured flux, only the NIRSpec+NIRISS measurements
  remain well-constrained.}
  \label{fig:albedo}
\end{figure}

Adopting the approaches of \citet{Schwartz2015} and \citet{Splinter2025}, we create a $\chi^2$ surface using our estimates for the dayside and nightside effective temperatures, and Equations 3 and 4 (see \autoref{fig:albedo}). We compute $\chi^2$ over a $2000 \times 2000$ grid in $A_B$–$\epsilon$ parameter space. The $1\sigma$, $2\sigma$, and $3\sigma$ confidence intervals relative to the minimum $\chi^2$ correspond to $\Delta \chi^2 < [1, 4, 9]$, respectively. We calculate $\chi^2$ with both standard errors and with uncertainties inflated by $1/\gamma$, where $\gamma$ represents the captured flux fraction.

We derive the $\chi^2$ surfaces using the dayside and nightside effective temperatures estimated with NIRSpec, NIRSpec+NIRISS and Spitzer. For Spitzer, brightness temperatures at $3.6\space \mu\mathrm{m}$ and $4.5\space \mu\mathrm{m}$ are combined using the Gaussian process regression to obtain $T_{\mathrm{day}}$ and $T_{\mathrm{night}}$ \citep{Dragomir2020, Crossfield2020}.

\subsubsection{Energy Balance Model}

To validate our estimated Bond albedo and extract the atmospheric properties of \planetname{}, we employ a semi-analytical energy balance model \texttt{Bell\_EBM} \citep{Cowan2011, Bell2018}. The model assumes the atmosphere consists of a single, fully mixed layer that absorbs all incoming stellar radiation. \texttt{Bell\_EBM} accounts for hydrogen dissociation and assumes constant-velocity zonal flows \citep{Bell2018}. This \texttt{Bell\_EBM} has been used in former phase curve analyses \citep[see][]{Mansfield2020,Dang2021,Davenport2025,Splinter2025}.

\begin{figure}[t]
  \centering
  \includegraphics[width=1.\linewidth]{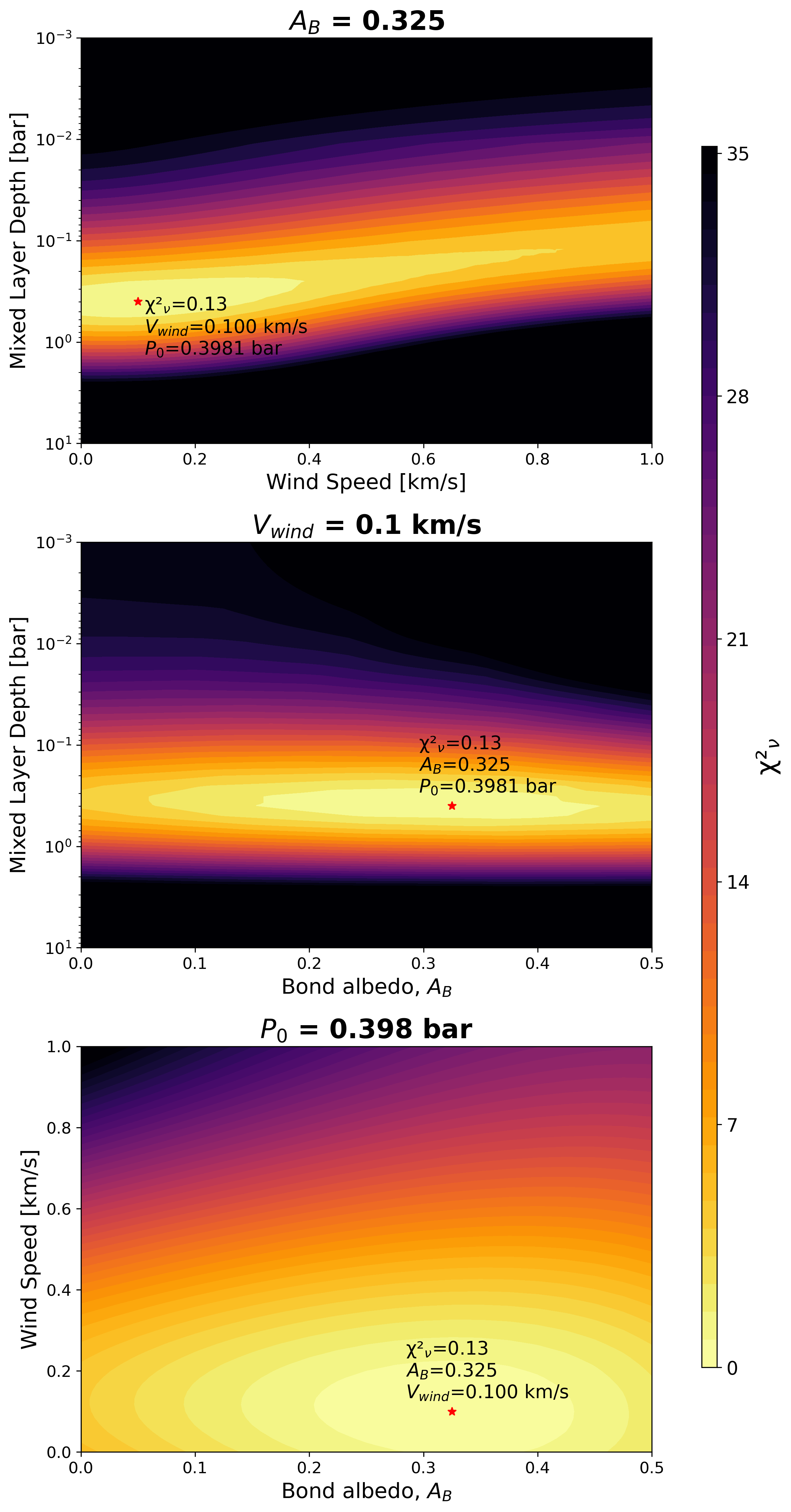}  % or .pdf, .jpg
  \caption{$\chi^2$ contour plots showing slices of the parameter space from fitting the NIRSpec effective temperature observations at 20 orbital phases (left panel of \autoref{fig:teff_appendix}) to the \texttt{Bell\_EBM} grid. The red stars are the locations of the best-fit model with reduced $\chi^2 = 0.13$. Each panel shows a 2D slice of the 3D parameter space, holding one parameter fixed at its best-fit value. This analysis predicts best-fit values of $A_B = 0.325$, $\mathrm{v}_{\mathrm{wind}} = 0.1$ $\mathrm{km/s}$ and $P_0 = 0.398$ $\mathrm{bar}$. The best-fit Bond albedo is similar to that inferred from the dayside and nightside temperatures.}
  \label{fig:ebm}
\end{figure}

We fit our phase-resolved effective temperatures to a grid of energy balance models. All orbital, planetary and stellar parameters are fixed to those in \citet{Jenkins2020} while the Bond albedo $A_B$, wind speed $\mathrm{v}_{\mathrm{wind}}$ and pressure at the bottom of the mixed layer, $P_0$, are varied. The mixed layer depth represents the pressure level at which the atmosphere responds to diurnal and seasonal forcing \citep{Dang2021}. The model assumes a uniform Bond albedo and constant wind speed across the planet. 

In our grid, we include 21 albedos $A_B\in[0,0.5]$, 61 wind speeds $\mathrm{v}_{\mathrm{wind}} \in [0,1]$ $\mathrm{km/s}$ and 41 logarithmically-spaced mixed layer depths $P_0 \in [0.001,10]$ $\mathrm{bar}$. \texttt{Bell\_EBM} also supports conversions between white-light curves and phase-resolved effective temperatures. Since thermal emission fits from NIRISS are only available at six phases, we strictly fit the EBM to the NIRSpec-derived effective temperatures, which sample 20 phases. We calculate the $\chi^2$ for each model and the contour plots of this fit are shown in \autoref{fig:ebm}.

\subsection{Energy Budget Constraints}

Capturing  $\sim$71\% -- 93\% of the planet's emitted flux, the combination of the NIRSpec/G395H and NIRISS/SOSS thermal phase curves provide the most precise constraints on the energy budget of \planetname{} or any exoplanet of its size to date. However, NIRSpec alone only captures $\sim$16 -- 38\% of the bolometric flux and is less robust in characterizing the energy budget of \planetname{}. 

Both the NIRSpec and NIRSpec+NIRISS dayside and nightside effective temperatures agree with the results from NIRISS in \citet{Coulombe2025}. The two nightside temperatures we find overlap at 2$\sigma$ and the addition of the NIRISS emission improves our constraints on the nightside. For instance, our Bond albedo estimate from NIRSpec alone is highly uncertain, resulting from the instrument's insensitivity to shorter wavelengths. The albedo becomes even more uncertain after inflating the errors in effective temperature by the inverse of the captured flux fraction \citep[see \autoref{fig:albedo} bottom panel;][]{Splinter2025}. Conversely, our estimate of $A_B = 0.29\pm 0.03$ from the NIRSpec+NIRISS data agrees well with Bond albedo estimates from NIRISS alone \citep{Coulombe2025}.

Our measured Bond albedo of \(A_B = 0.29 \pm 0.03\) is notably lower than the geometric albedo of $A_g = 0.50\pm0.07$ reported by \citet{Coulombe2025}, indicating that \planetname's reflection is dominated by backscattering rather than uniform (Lambertian) scattering. 

\subsection{Atmospheric Heat Transport}

The planet's low heat recirculation efficiency $(\epsilon\sim0.15)$ implies limited atmospheric heat transport between the dayside and nightside. This thermal asymmetry could play a key role in shaping the longitudinal variation of chemical abundances.

By fitting a grid of \texttt{Bell\_EBM} models to our phase-resolved effective temperatures from NIRSpec, the best-fit model prefers a slightly higher Bond albedo of $A_B = 0.325$. This inflation is also noted in \citet{Splinter2025}. Our \texttt{Bell\_EBM} fit also suggests low zonal wind speeds of 0.1 $\mathrm{km/s}$. Such weak winds are consistent with the planet’s inefficient day-night heat redistribution and can suppress horizontal mixing. These slow winds could result from magnetic drag if the atmosphere is coupled to a planetary magnetic field \citep{Perna2010}. Given \planetname{}’s high temperatures, ionization of Na and K, and consequent coupling to the planetary magnetic field, are likely. Future phase curve analyses and circulation models should investigate the possible role of magnetic effects in shaping \planetname{}’s climate.

\section{Conclusions} 
\label{sec:conclusion}

The phase-resolved G395H emission spectra of \planetname{} depicts an ultra-hot Neptune with a highly enriched atmosphere exhibiting remarkable day–night contrasts. The dayside emission is marked by strong CO, CO$_2$ and H$_2$O absorption from a hot, relatively cloud-free atmosphere, whereas the nightside is blanketed by silicate clouds that mute the H$_2$O signature and produce a blackbody-like spectrum. The tentative identification of SO$_2$ on the western nightside, if confirmed, would add \planetname{} as the first ultra-hot Neptune to the growing list of highly-irradiated gas planets with active photochemistry. 

%These findings point to an atmosphere rich in heavy elements (high metallicity) with a near-solar C/O ratio, consistent with formation beyond the ice line followed by inward migration. \planetname’s reflective dayside clouds, dynamic species abundances, and possible photochemistry provide a unique glimpse into the climate of an extreme Neptune-mass planet. Understanding this unyielding planet's survival amid the hot Neptune desert helps us appreciate the universe's most resilient creations, and underscores the power of JWST phase-curve observations in mapping exotic worlds in multiple dimensions.

\planetname{}'s super-Solar metallicity and likely high C/O, along with initial indications of atmospheric sulfur, may help us constrain its formation and migratory history. Atmospheric C/O has been put forward as a tracer of planet formation location \citep{Oberg2011}. C/O~$\sim1$ would then be evidence of gas accretion outside the water snowline, and potentially even $\sim10$~AU away between the CO$_2$ and CO snowlines, with subsequent significant migration. Atmospheric SO$_2$, if further confirmed and more precisely measured with mid-infrared observations, could also inform the types of accretion (pebble for super-Solar S/O vs. planetesimal for Solar/sub-Solar S/O) that \planetname{} may have undergone in its early existence. However, without a detailed understanding of the chemistry in \planetname{}'s natal disk, or the true distribution of elements in its atmosphere, and what sort of mixing and other disequilibrium processes may be ongoing within it, we can only hint at its history. 

%\acknowledgments

% \textbf{Acknowledgments}  

%We would also like to thank our anonymous reviewers for their constructive comments and suggestions.

% \software{\texttt{PySynPhot} \citep{pysynphot}; \texttt{MultiNest} \citep{Feroz2009}; \poseidon \citep{MacDonald2017, MacDonald2023}; \texttt{PyMultiNest} \citep{Buchner2014}}.

%% To help institutions obtain information on the effectiveness of their 
%% telescopes the AAS Journals has created a group of keywords for telescope 
%% facilities.
%
%% Following the acknowledgments section, use the following syntax and the
%% \facility{} or \facilities{} macros to list the keywords of facilities used 
%% in the research for the paper.  Each keyword is check against the master 
%% list during copy editing.  Individual instruments can be provided in 
%% parentheses, after the keyword, but they are not verified.

%\vspace{5mm}
% \facilities{JWST(NIRSpec), Exo.Mast, NASA ADS}

%% Similar to \facility{}, there is the optional \software command to allow 
%% authors a place to specify which programs were used during the creation of 
%% the manusscript. Authors should list each code and include either a
%% citation or url to the code inside ()s when available.

%\software{astropy \citep{2013A&A...558A..33A}, Cloudy \citep{2013RMxAA..49..137F}, SExtractor \citep{1996A&AS..117..393B}}

\section*{Acknowledgements}
The authors would like to thank Jacob Lustig-Yaeger and Giannina Guzman for their contributions to this work. 

This research is based on observations made with the NASA/ESA James Webb Space Telescope obtained from the Space Telescope Science Institute (STScI), which is operated by the Association of Universities for Research in Astronomy, Inc., under NASA contract JWST-GO-03231.001-A. 

This research has made use of NASA's Astrophysics Data System Bibliographic Services and the NASA Exoplanet Archive \citep{exoplanetArchive_PS}, which is operated by the California Institute of Technology, under contract with NASA under the Exoplanet Exploration Program.

Some of the data presented in this article was obtained from the Mikulski Archive for Space Telescopes (MAST) at the STScI. The specific observations analyzed can be accessed via \dataset[DOI]{https://doi.org/10.17909/0nxz-sz76}.

N.B.C. acknowledges support from an NSERC Discovery Grant, a Tier 2 Canada Research Chair, and an Arthur B.\ McDonald Fellowship and thanks the Trottier Space Institute and l'Institut de recherche sur les exoplanètes for their financial support and dynamic intellectual environment.

O.V. acknowledges funding from Agence Nationale de la Recherche (ANR), project ``EXACT'' (ANR-21-CE49-0008-01) as well as from the Centre National d’\'{E}tudes Spatiales (CNES).

JSJ greatfully acknowledges support by FONDECYT grant 1240738 and from the ANID BASAL project FB210003.

% \facilities{\citealt{exoplanetArchive_PS}} 

% \software{Python}}

\onecolumngrid            % switch to single column for the whole appendix
\clearpage
\appendix

\label{appendix}
\renewcommand{\thetable}{A\arabic{table}}
\renewcommand{\thefigure}{A\arabic{figure}}
\setcounter{table}{0}
\setcounter{figure}{0}

\section{Additional Data Analysis Materials}
\label{appendix:more_data_analysis}

This appendix provides supplementary data analysis materials to support the results presented in the main text. The figures and tables included here illustrate intermediate steps in the data reduction process and demonstrate the robustness of the adopted modeling choices.

\begin{figure*}[b]
    \centering
    \includegraphics[width=0.9\linewidth, trim={0cm 0cm 0cm 0cm}, clip]{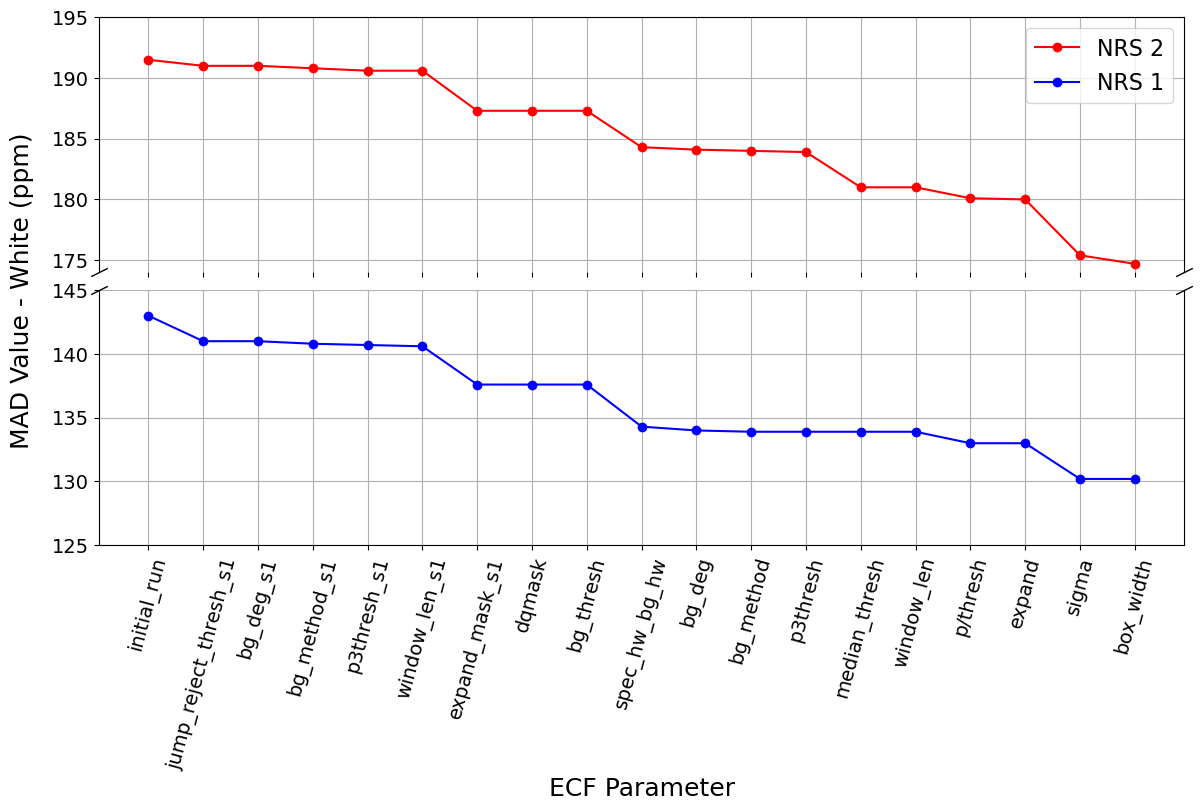}
    \caption{Optimization histories for the parameter-by-parameter data reductions with the \texttt{Eureka!} pipeline.}
    \label{fig:optimization}
\end{figure*}

\begin{deluxetable*}{llrr}
\tablewidth{0pt}
\tablecaption{Best-fit stellar and orbital parameters with uncertainties. \label{tab:fit_parameters}}
\tablehead{
Parameter & Unit & \texttt{Eureka! (v1)} & \texttt{ExoTEDRF\hspace{0.05cm}+\hspace{0.05cm}Eureka! (v2)}
}
\startdata
F$_p$/F$_s$           & \nodata  & $0.000453 \pm 0.0000102$ & $0.000452 \pm 0.0000104$ \\
R$_p$/R$_s$           & \nodata  & $0.04615 \pm 0.000611$   & $0.04609 \pm 0.000429$ \\
Period, $P$           & days     & $0.792064 \pm 1.04\times10^{-7}$ & $0.792064 \pm 1.04\times10^{-7}$ \\
$t_0$ (MJD offset)    & MJD      & $0.80039 \pm 0.000058$   & $0.80057 \pm 0.000062$ \\
$t_{\rm sec}$         & MJD      & $0.40436 \pm 0.000152$   & $0.40440 \pm 0.000150$ \\
Inclination, $i$      & degrees  & $76.72 \pm 0.223$        & $76.81 \pm 0.096$ \\
$a/R_s$               & \nodata  & $3.960 \pm 0.0537$       & $3.981 \pm 0.0237$ \\
Eccentricity, $e$     & \nodata  & $0$                      & $0$ \\
Argument of periastron, $\omega$ & degrees & $90$           & $90$ \\
\hline
$u_1$                 & \nodata  & $0.075 \pm 0.086$        & $0.102 \pm 0.074$ \\
$u_2$                 & \nodata  & $0.118 \pm 0.084$        & $0.207 \pm 0.069$ \\
\enddata
\tablecomments{
Uncertainties are symmetric, computed as the mean of the reported $-1\sigma$ and $+1\sigma$ bounds. 
$v1$ values come from Eureka! alone; $v2$ values from ExoTEDRF+Eureka.
}
\end{deluxetable*}

\begin{figure*}[t]
  \centering
  \begin{subfigure}{0.49\textwidth}
    \centering
    \includegraphics[width=\linewidth]{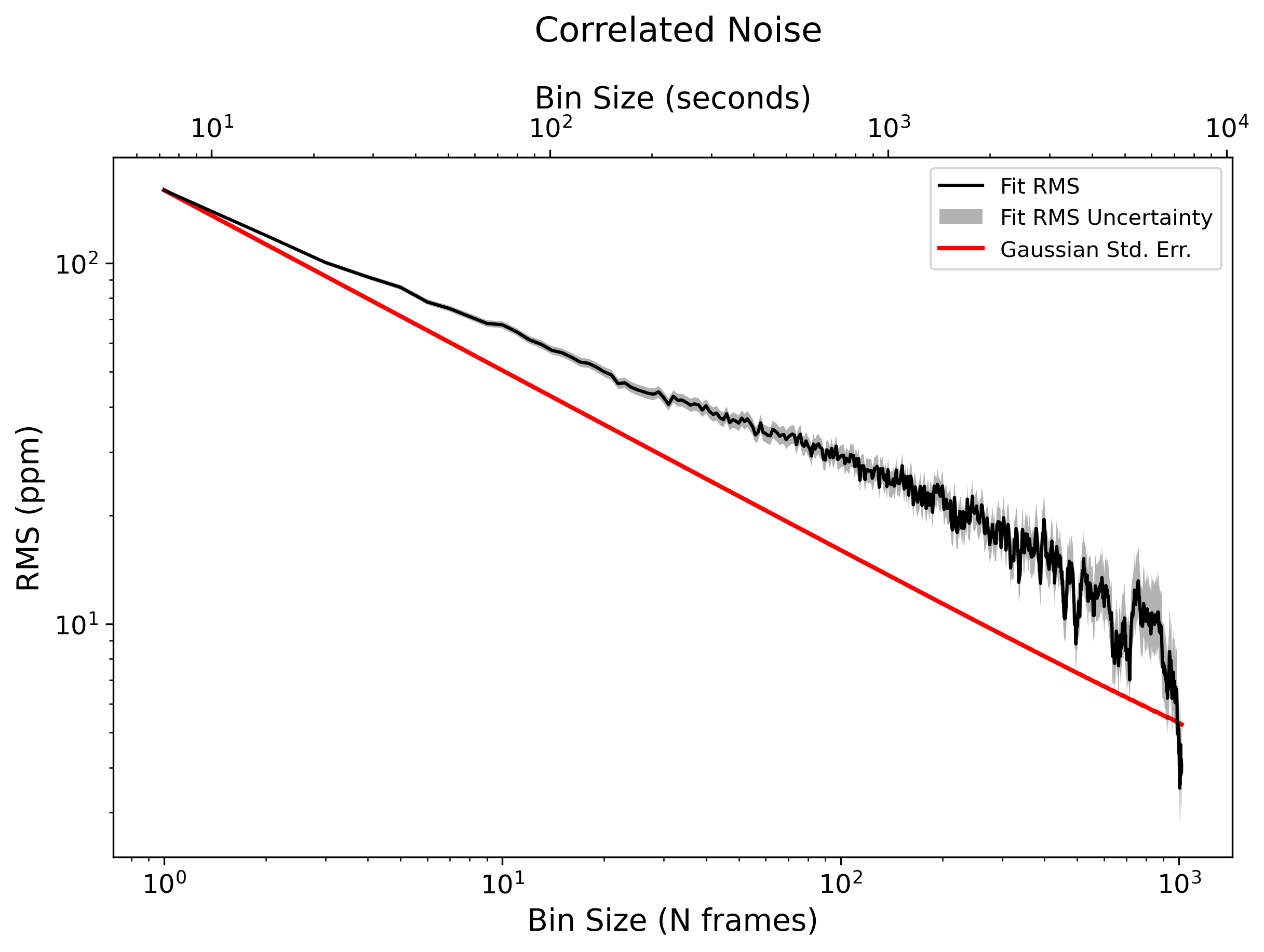} 
    \subcaption{NRS1}
    \label{fig:allan_nrs1}
  \end{subfigure}\hfill
  \begin{subfigure}{0.49\textwidth}
    \centering
    \includegraphics[width=\linewidth]{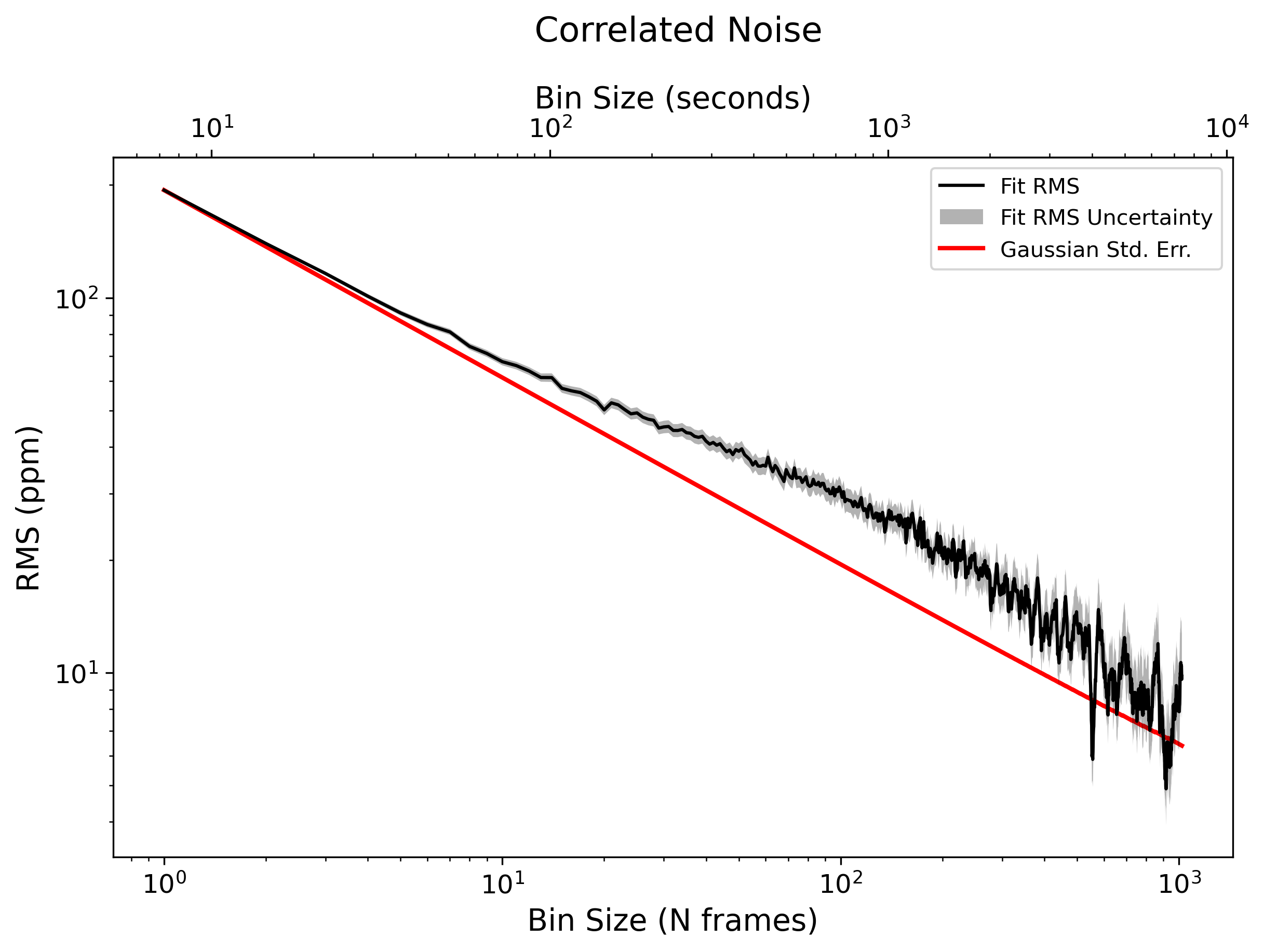}
    \subcaption{NRS2}
    \label{fig:allan_nrs2}
  \end{subfigure}
  \caption{
  Correlated-noise diagnostics for the white light curve residuals for NRS1 and NRS2. The black curve traces the fitted RMS of binned residuals with a shaded $1\sigma$ envelope; the red line marks the Gaussian expectation ($\mathrm{RMS}\!\propto\!N^{-1/2}$), i.e., the “Allan limit”. Consistency of the black curve with the red slope indicates predominantly white noise; sustained excess above the line reveals time-correlated (red) noise. Both channels closely follow the $N^{-1/2}$ trend across the binning range relevant for transit and eclipse timescales, with NRS2 showing slightly tighter conformity at large bins. These tests confirm that residual systematics are well controlled and that the final white-light solutions are not limited by red noise.
  }
  \label{fig:correlated_noise}
\end{figure*}

%MOVE These 2 TO APPENDIX? This doesnt really add much
\begin{figure*}[]
    \centering
    \includegraphics[width=0.9\linewidth, trim={0cm 0cm 0cm 0cm}, clip]{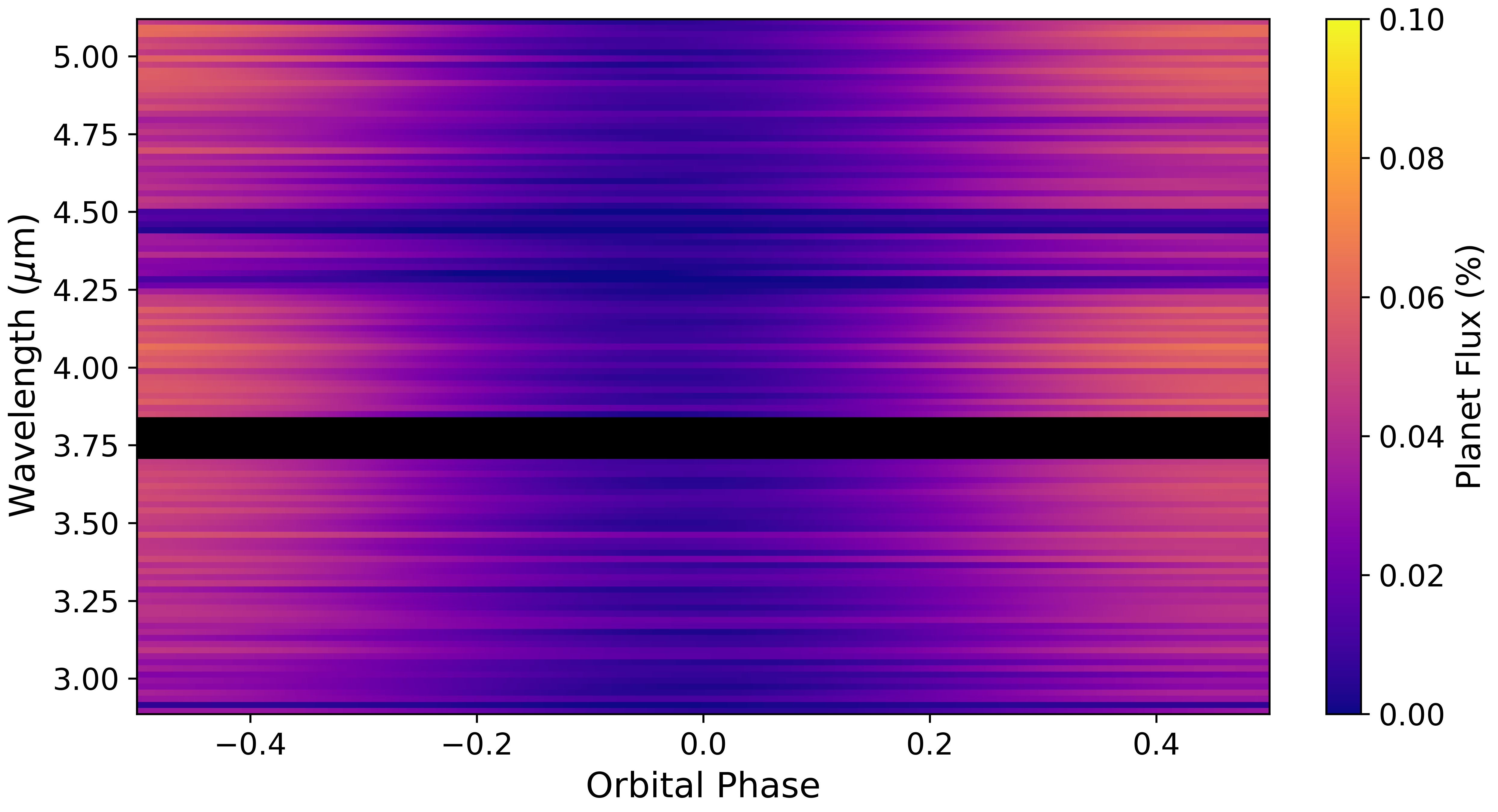}
    \caption{2-D emission spectrum for the \eureka dataset (v1).}
    \label{fig:2D_Spectrum}
\end{figure*}

\begin{figure*}[]
    \centering
    \includegraphics[width=0.9\linewidth, trim={0cm 0cm 0cm 0cm}, clip]{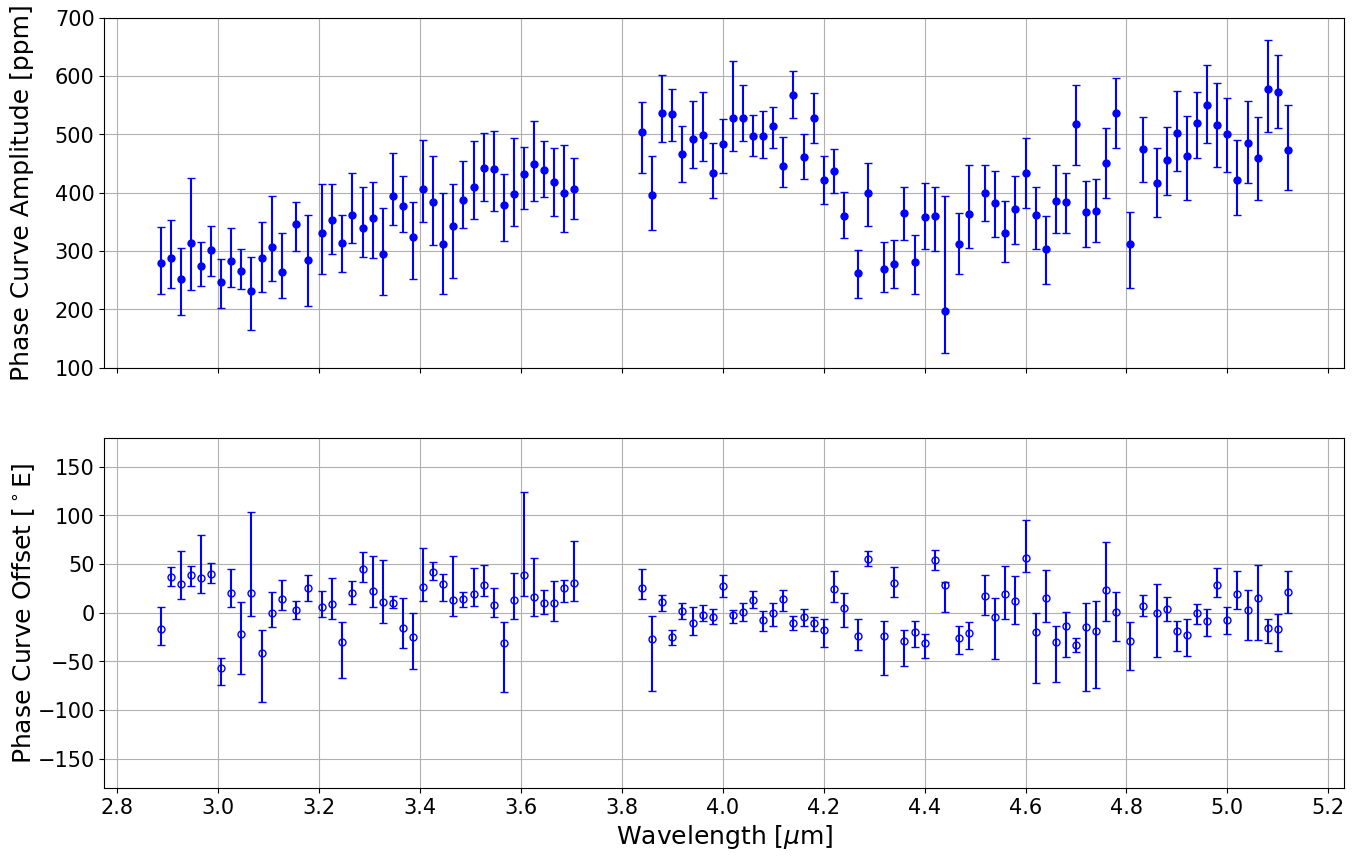}
    \caption{Phase curve amplitude and offset for the \eureka dataset (v1).}
    \label{fig:PC_Amp_Off}
\end{figure*}

\begin{figure*}[t]
    \centering
    \includegraphics[width=1.0\linewidth, trim={0cm 0cm 0cm 0cm}, clip]{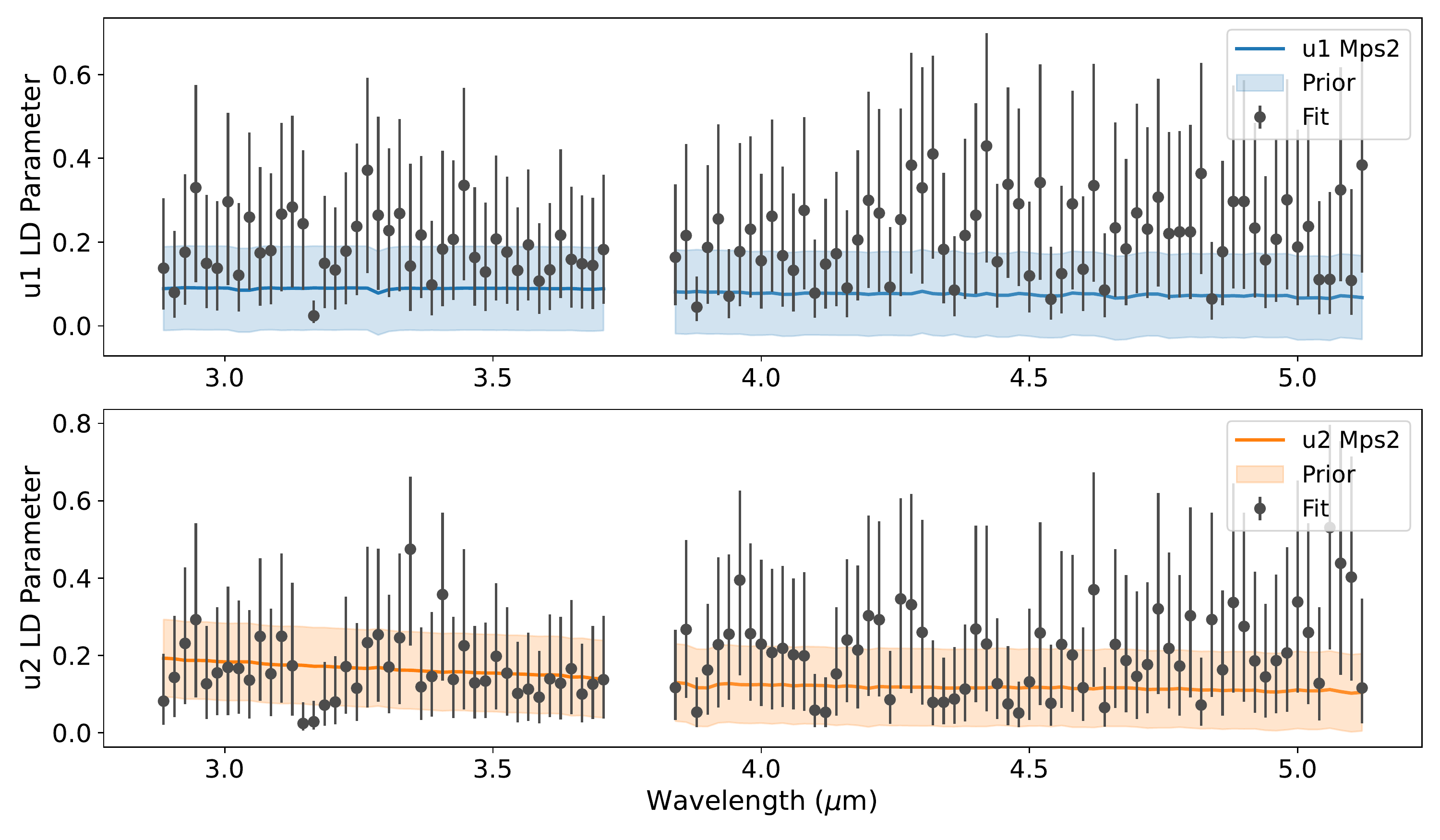}
    \caption{Comparison of the \texttt{exotic-ld} mps2 stellar model's predicted $u1$ and $u2$ limb darkening coefficients and priors, against the free-fitted white light curve parameters for the dayside emission spectrum of the Eureka! data reduction (v1).}
    \label{fig:ld_comparison}
\end{figure*}

\begin{figure*}[t]
    \centering
    \includegraphics[width=1.0\linewidth, trim={0cm 0cm 0cm 0cm}, clip]{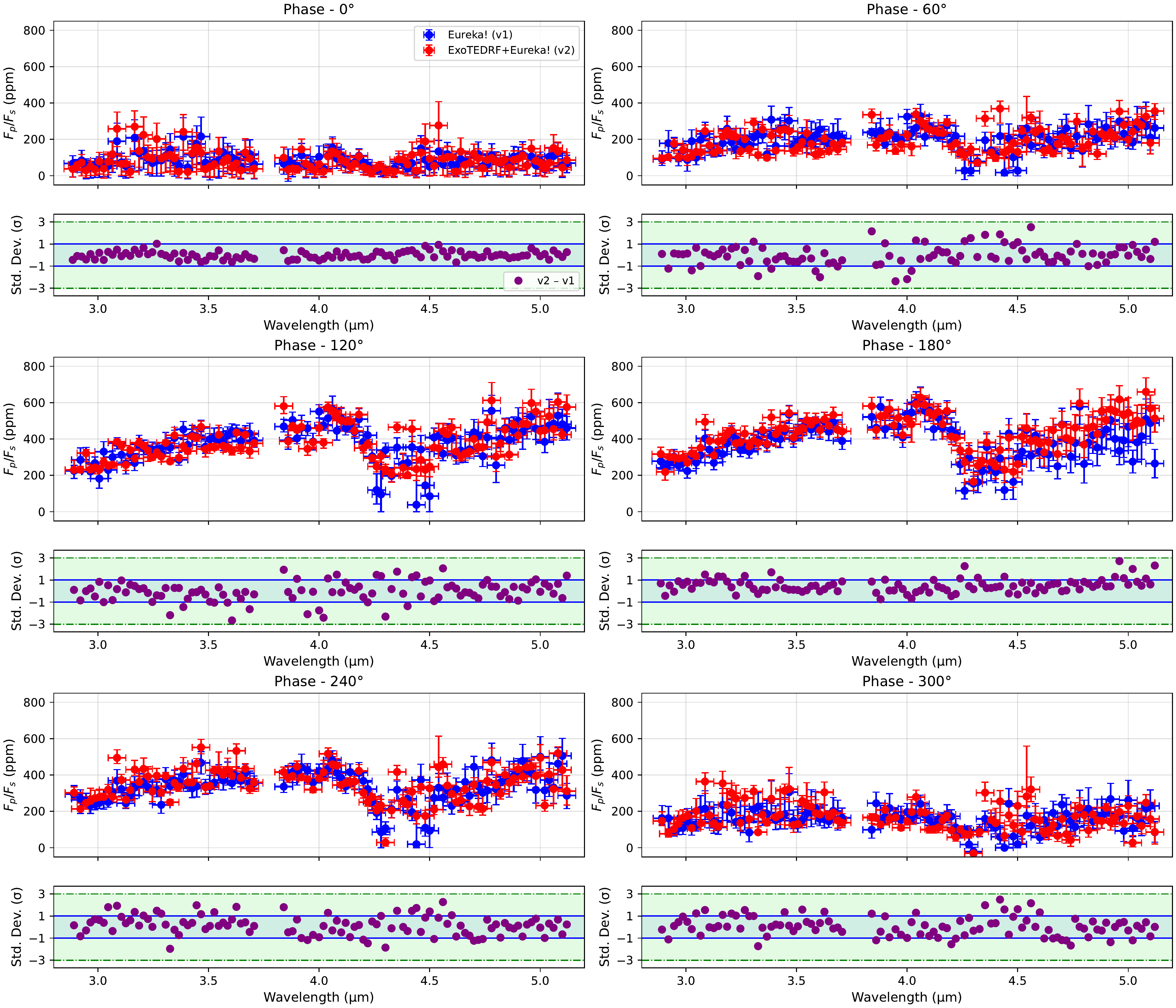}
    \caption{Comparison of the phase-resolved emission spectra for the Eureka! data reduction (v1) and the ExoTEDRF+Eureka! reduction (v2).}
    \label{fig:reduction comparison}
\end{figure*}

\clearpage
\onecolumngrid      % switch to single column for the whole appendix
\section{Additional Retrieval Materials} 
\label{appendix:more_retrievals}
\renewcommand{\thetable}{B\arabic{table}}
\renewcommand{\thefigure}{B\arabic{figure}}
\setcounter{table}{0}
\setcounter{figure}{0}

This appendix contains supplementary retrieval results across multiple orbital phases. These additional tables highlight the relative performance of different atmospheric models, providing further context to the retrieval comparisons discussed in the main body of the paper.

\begin{table*}[ht]
\centering
\caption{Retrieval comparisons across orbital phases.}
\label{tab:significances}
\begin{tabular}{llrrr}
\toprule
 & Retrieval Details & ln(Z) & ln(B) & Rejection $\sigma$ \\
\midrule
\multicolumn{5}{l}{\textbf{Phase - 0$^\circ$}} \\
Model 1 & Full Model & 916.51 & 0.00 & - \\
Model 2 & No CO      & 914.66 & 1.85 & 2.47 \\
Model 3 & No CO2     & 913.53 & 2.98 & 2.94 \\
Model 4 & No H2O     & 916.83 & -0.32 & - \\
Model 5 & No SO2     & 916.47 & 0.04 & 1.10 \\
\midrule
\multicolumn{5}{l}{\textbf{Phase - 60$^\circ$}} \\
Model 1 & Full Model & 892.65 & 0.00 & - \\
Model 2 & No CO      & 888.70 & 3.95 & 3.28 \\
Model 3 & No CO2     & 876.01 & 16.64 & 6.10 \\
Model 4 & No H2O     & 893.60 & -0.95 & - \\
Model 5 & No SO2     & 893.63 & -0.98 & - \\
\midrule
\multicolumn{5}{l}{\textbf{Phase - 120$^\circ$}} \\
Model 1 & Full Model & 874.60 & 0.00 & - \\
Model 2 & No CO      & 870.57 & 4.03 & 3.31 \\
Model 3 & No CO2     & 847.27 & 27.33 & 7.68 \\
Model 4 & No H2O     & 875.05 & -0.45 & - \\
Model 5 & No SO2     & 877.48 & -2.88 & - \\
\midrule
\multicolumn{5}{l}{\textbf{Phase - 180$^\circ$}} \\
Model 1 & Full Model & 861.33 & 0.00 & - \\
Model 2 & No CO      & 840.62 & 20.71 & 6.75 \\
Model 3 & No CO2     & 796.98 & 64.35 & 11.57 \\
Model 4 & No H2O     & 855.26 & 6.07 & 3.91 \\
Model 5 & No SO2     & 862.72 & -1.39 & - \\
\midrule
\multicolumn{5}{l}{\textbf{Phase - 240$^\circ$}} \\
Model 1 & Full Model & 881.98 & 0.00 & - \\
Model 2 & No CO      & 873.61 & 8.37 & 4.49 \\
Model 3 & No CO2     & 845.66 & 36.32 & 8.79 \\
Model 4 & No H2O     & 880.91 & 1.07 & 2.07 \\
Model 5 & No SO2     & 881.77 & 0.21 & 1.38 \\
\midrule
\multicolumn{5}{l}{\textbf{Phase - 300$^\circ$}} \\
Model 1 & Full Model & 880.65 & 0.00 & - \\
Model 2 & No CO      & 874.75 & 5.90 & 3.87 \\
Model 3 & No CO2     & 864.82 & 15.83 & 5.96 \\
Model 4 & No H2O     & 881.15 & -0.50 & - \\
Model 5 & No SO2     & 878.82 & 1.83 & 2.46 \\
\bottomrule
\end{tabular}
\end{table*}

\begin{figure*}[t]
    \centering
    \includegraphics[width=1.0\linewidth, trim={0cm 0cm 0cm 0cm}, clip]{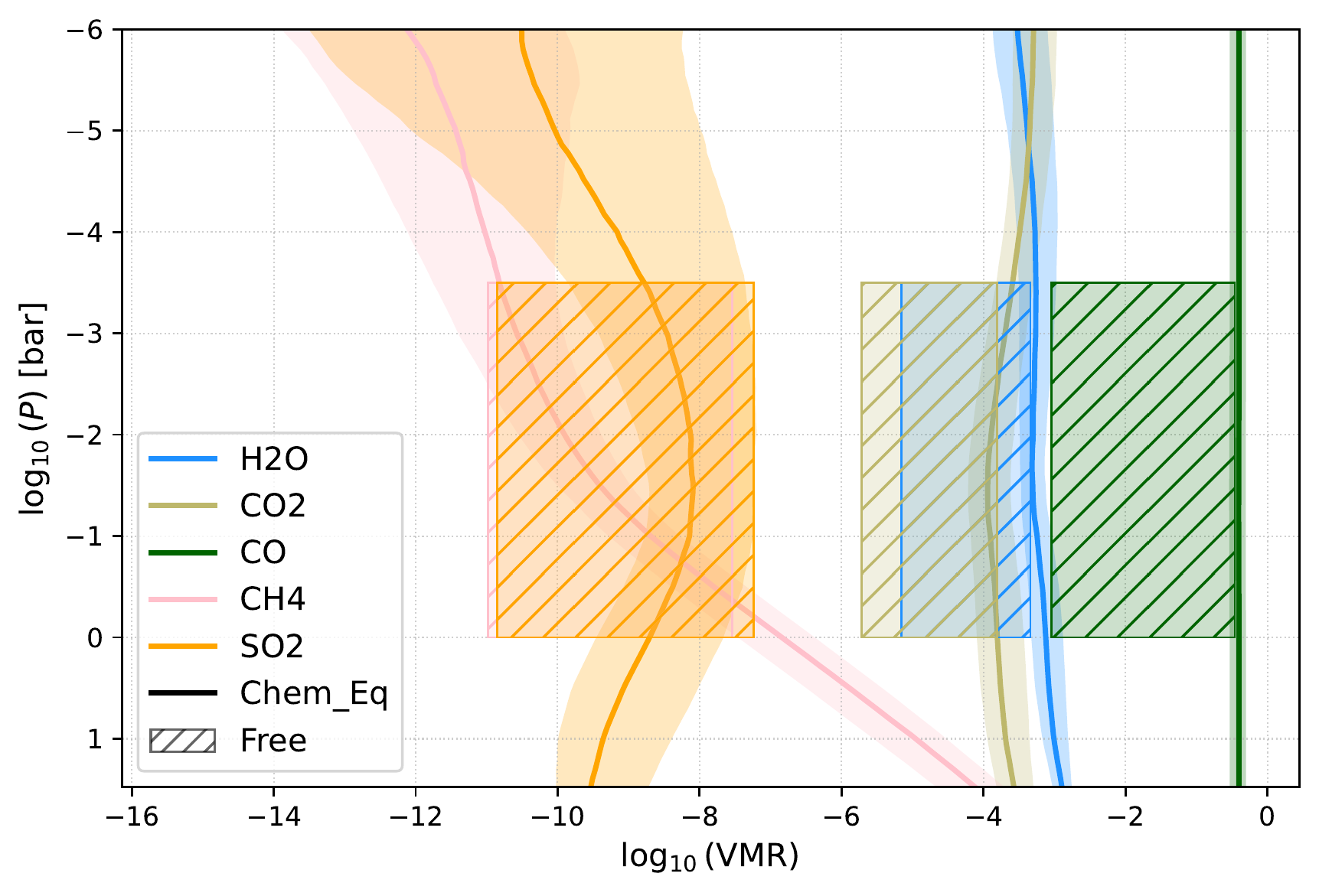}
    \caption{
    Comparison of free-retrieved and chemical–equilibrium volume mixing ratios (VMR's) for \planetname{} across the photospheric pressure range probed in emission (0 to $-3.5$ in $\log_{10}(P\,[\mathrm{bar}])$; see \autoref{fig:Pressure_Contribution}). Ribboned regions show the free retrieval $1\sigma$ ranges for H$_2$O, CO$_2$, CO, CH$_4$, and SO$_2$; solid curves show the corresponding chemical–equilibrium profiles. CO$_2$ exhibits close agreement between approaches over the full pressure interval. For CO and H$_2$O, the equilibrium solutions track the upper edge of the free posteriors, supporting the interpretation that median free values are effectively upper limits set by the data. CH$_4$ and SO$_2$ remain low in both frameworks.
    }
    \label{fig:VMR_comparison_free_chemeq}
\end{figure*}

\clearpage
\section{Additional Energy Budget Materials} 
\label{appendix:more_energy_budget}
\renewcommand{\thetable}{C\arabic{table}}
\renewcommand{\thefigure}{C\arabic{figure}}
\setcounter{table}{0}
\setcounter{figure}{0}

This appendix presents extended results from the planet’s energy budget analysis. The figures included here show alternative effective temperature estimates derived from different binning and dataset combinations, complementing the phase-resolved results in the main text.

\begin{figure*}[b]
  \centering
  \includegraphics[width=0.95\linewidth]{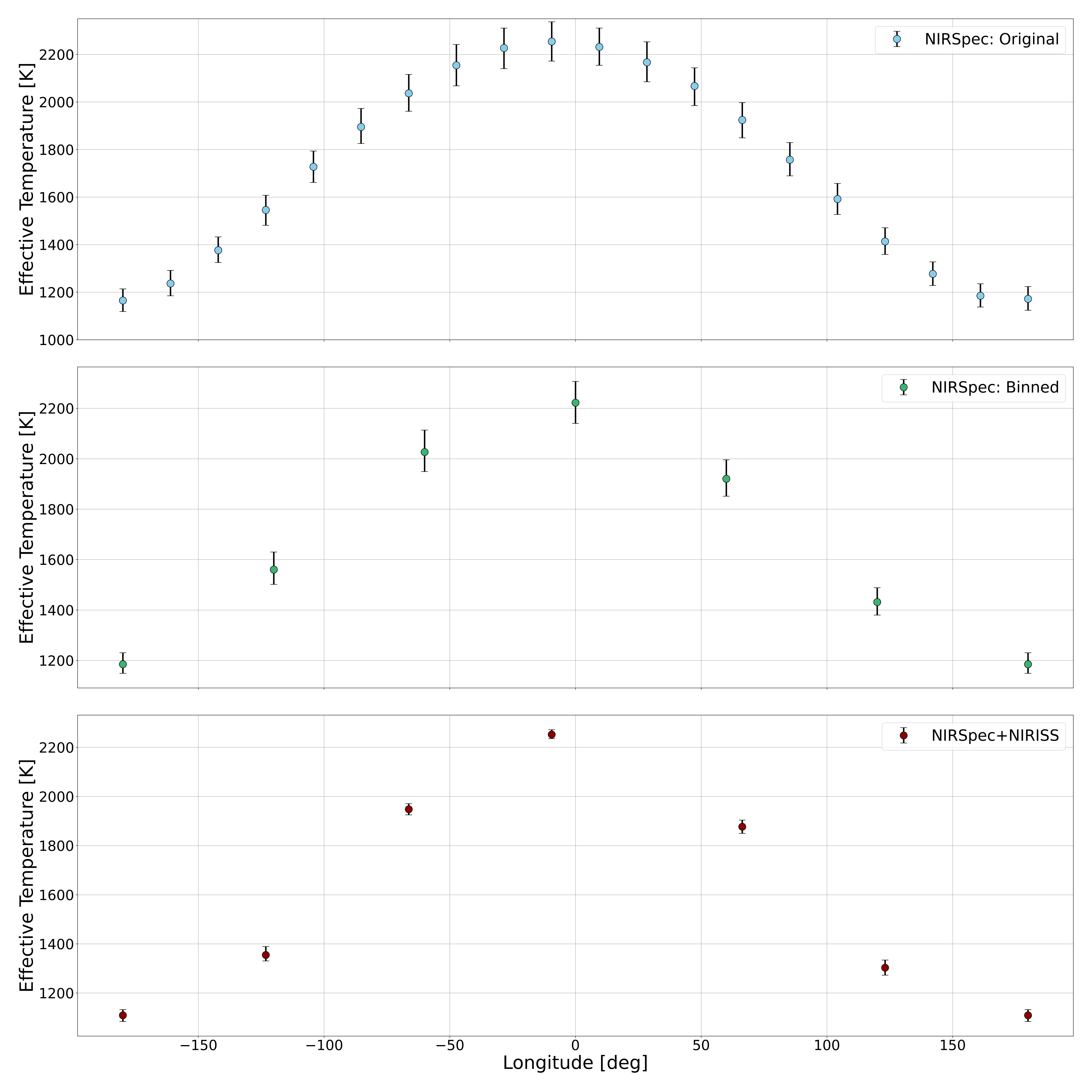}  % or .pdf, .jpg
  \caption{Longitudinally-resolved effective temperature estimates of \planetname{} using the Gaussian Process outlined in \citet{Pass2019}. \textit{Top:} Effective temperatures estimated with the NIRSpec/G395H phase curve. We report the dayside and nightside temperatures based on the full-resolution data, prior to any binning.  \textit{Middle:}  Effective temperature estimates with binned NIRSpec/G395H data. We divided the phase curve into six equal-width bins of 30° in orbital phase, centered at evenly spaced phases. For each bin and wavelength, the flux was taken as the arithmetic mean of all measurements within the bin. Lower and upper uncertainties were propagated by combining the individual errors in quadrature and scaling by the number of points in the bin. \textit{Bottom:} Effective temperature estimates with the combined NIRSpec+NIRISS phase curve. Temperatures are calculated by combining the binned NIRSpec and median-fit NIRISS flux measurements.
  }
  \label{fig:teff_appendix}
\end{figure*}

% \newpage

\clearpage            % stronger than \newpage: flushes floats
\FloatBarrier         % belt-and-suspenders: stop any remaining float drift
\bibliography{main}
\end{document}

%% file: kbsmacs.tex
\newcommand\degree{\degr}
\newcommand\degrees\degree

\DeclareSymbolFont{UPM}{U}{eur}{m}{n}
\DeclareMathSymbol{\umu}{0}{UPM}{"16}
\let\oldumu=\umu
\renewcommand\umu{\ifmmode\oldumu\else\math{\oldumu}\fi}

\newcommand\microns \micron

\let\oldsim=\sim
\renewcommand\sim{\ifmmode\oldsim\else\math{\oldsim}\fi}
\let\oldpm=\pm
\renewcommand\pm{\ifmmode\oldpm\else\math{\oldpm}\fi}
\newcommand\by{\ifmmode\times\else\math{\times}\fi}

\newbox{\wdbox}
\renewcommand\c{\setbox\wdbox=\hbox{,}\hspace{\wd\wdbox}}
\renewcommand\i{\setbox\wdbox=\hbox{i}\hspace{\wd\wdbox}}
\newcommand\n{\hspace{0.5em}}

\newcount\timect
\newcount\hourct
\newcount\minct
\newcommand\now{\timect=\time \divide\timect by 60
         \hourct=\timect \multiply\hourct by 60
         \minct=\time \advance\minct by -\hourct
         \number\timect:\ifnum \minct < 10 0\fi\number\minct}

\catcode`@=11

\newcommand\comment[1]{}
\newcommand\commenton{\catcode`\%=14}
\newcommand\commentoff{\catcode`\%=12}

\renewcommand\math[1]{$#1$}
\newcommand\mathshifton{\catcode`\$=3}
\newcommand\mathshiftoff{\catcode`\$=12}

\comment{the backslash is necessary}

\comment{alignment tab}

\let\atab=&
\newcommand\atabon{\catcode`\&=4}
\newcommand\ataboff{\catcode`\&=12}

\let\oldmsp=\sp
\let\oldmsb=\sb
\def\sp#1{\ifmmode
           \oldmsp{#1}%
         \else\strut\raise.85ex\hbox{\scriptsize #1}\fi}
\def\sb#1{\ifmmode
           \oldmsb{#1}%
         \else\strut\raise-.54ex\hbox{\scriptsize #1}\fi}
\newbox\@sp
\newbox\@sb
\def\sbp#1#2{\ifmmode%
           \oldmsb{#1}\oldmsp{#2}%
         \else
           \setbox\@sb=\hbox{\sb{#1}}%
           \setbox\@sp=\hbox{\sp{#2}}%
           \rlap{\copy\@sb}\copy\@sp
           \ifdim \wd\@sb >\wd\@sp
             \hskip -\wd\@sp \hskip \wd\@sb
           \fi
        \fi}
\def\msp#1{\ifmmode
           \oldmsp{#1}
         \else \math{\oldmsp{#1}}\fi}
\def\msb#1{\ifmmode
           \oldmsb{#1}
         \else \math{\oldmsb{#1}}\fi}
\def\supon{\catcode`\^=7}
\def\supoff{\catcode`\^=12}
\def\subon{\catcode`\_=8}
\def\suboff{\catcode`\_=12}
\def\supsubon{\supon \subon}
\def\supsuboff{\supoff \suboff}

\newcommand\actcharon{\catcode`\~=13}
\newcommand\actcharoff{\catcode`\~=12}

\newcommand\paramon{\catcode`\#=6}
\newcommand\paramoff{\catcode`\#=12}

\comment{And now to turn us totally on and off...}

\newcommand\reservedcharson{\commenton \mathshifton \atabon \supsubon \actcharon
	\paramon}

\newcommand\reservedcharsoff{\commentoff \mathshiftoff \ataboff
	\supsuboff \actcharoff \paramoff}

\catcode`@=12
\reservedcharsoff

\reservedcharson

\comment{ Must have ONLY ONE of these... trust these macros, they work

}

\newcommand{\squishlist}{
 \begin{list}{$\bullet$}
  { \setlength{\itemsep}{0pt}
     \setlength{\parsep}{0pt}
     \setlength{\topsep}{0pt}
     \setlength{\partopsep}{0pt}
     \setlength{\leftmargin}{2.0em}
     \setlength{\labelwidth}{1.5em}
     \setlength{\labelsep}{0.5em} } }

\newcommand{\squishlisttwo}{
 \begin{list}{$\bullet$}
  { \setlength{\itemsep}{1pt}
     \setlength{\parsep}{3pt}
     \setlength{\topsep}{3pt}
     \setlength{\partopsep}{0pt}
     \setlength{\leftmargin}{2.0em}
     \setlength{\labelwidth}{1.5em}
     \setlength{\labelsep}{0.5em} } }

\newcommand{\squishend}{
  \end{list}  }

\reservedcharson
\actcharon